\documentclass[useAMS,usenatbib]{mn2e}
\usepackage{epsfig}

\def\gsim{\;\rlap{\lower 2.5pt\hbox{$\sim$}}\raise 1.5pt\hbox{$>$}\;}
\def\lsim{\;\rlap{\lower 2.5pt\hbox{$\sim$}}\raise 1.5pt\hbox{$<$}\;}
\def\dML{\mbox{$\nabla_{\ell} \Upsilon$}}

\renewcommand{\arcsec}{\mbox{$^{\prime\prime}$}}

\newcommand{\degree}{\mbox{$^\circ$}}
\newcommand{\kms}{\mbox{\,km~s$^{-1}$}}

\renewcommand{\mag}{\mbox{\,mag}}

\newcommand{\PNS}{PN.S}

\def\N4374{\mbox{NGC~4374}}

\newcommand{\VD}{\mbox{\rm VD}}

\def\Re{\mbox{$\,R_{\rm e}$}}

\newcommand{\ML}{M/L}
\newcommand{\LCDM}{\mbox{$\Lambda$CDM}}

\def\vir{\mbox{$_{\rm vir}$}}

\def\Mvir{\mbox{$M_{\rm vir}$}}

\def\mag{\mbox{\,\rm mag}}
\def\Msun{\mbox{$M_\odot$}}

\def\Ystar{\mbox{$\Upsilon_*$}}

\def\Ysol{\mbox{$\Upsilon_{\odot, V}$}}

\def\Rm{\mbox{$R$}}
\def\vc{\mbox{$v_{\rm c}$}}

\title[Dark Matter in NGC~4374 with PN.S]{The PN.S Elliptical Galaxy Survey:
a standard $\Lambda$CDM halo around NGC~4374?\thanks{Based on
observations made with the William Herschel Telescope operated on
the island of La Palma by the Isaac Newton Group in the Spanish
Observatorio del Roque de los Muchachos of the Instituto de
Astrofisica de Canarias and ESO-VLT at the Cerro Paranal
(Chile).}} \vspace{-2cm}
\author[Napolitano et al.]{\noindent
N.R.~Napolitano$^{1}$\thanks{E-mail: napolita@na.astro.it (NRN);
romanow@ucolick.org (AJR)}, A.J.~Romanowsky$^{2,3}$,
M.~Capaccioli$^{4,5}$, N.G.~Douglas$^{6}$,\and
M.~Arnaboldi$^{7,8},$ L.~Coccato$^{9}$,
O.~Gerhard$^{9}$, K.~Kuijken$^{10}$, M.R.~Merrifield$^{11}$,\and
S.P. Bamford$^{11}$, A. Cortesi$^{11}$, P.~Das$^{9}$,
K.C.~Freeman$^{12}$
\\~\\
$^1$ INAF-Observatory of Capodimonte, Salita Moiariello, 16, 80131, Naples, Italy\\
$^2$ UCO/Lick Observatory, University of California, Santa Cruz, CA 95064, USA\\
$^3$ Departamento de F\'{i}sica, Universidad de Concepci\'{o}n, Casilla 160-C, Concepci\'on, Chile\\
$^4$ Dipartimento di Scienze Fisiche, Universit`a Federico II, Via Cinthia, 80126, Naples, Italy \\
$^5$ MECENAS, University of Naples Federico II and University of Bari, Italy\\
$^6$ Kapteyn Astronomical Institute, Postbus 800, 9700 AV Groningen, The Netherlands\\
$^7$ European Southern Observatory, Karl-Schwarzschild-Strasse 2,
D-85748 Garching, Germany\\
$^{8}$ INAF, Osservatorio Astronomico di Pino Torinese, I-10025
Pino
Torinese, Italy\\
$^{9}$ Max-Planck-Institut f\"ur Extraterrestriche Physik, Giessenbachstrasse, D-85748 Garching b. M\"unchen, Germany\\
$^{10}$ Leiden Observatory, Leiden University, PO Box 9513, 2300RA Leiden, The Netherlands \\
$^{11}$ School of Physics and Astronomy, University of Nottingham, University Park, Nottingham NG7 2RD, UK\\
$^{12}$ Research School of Astronomy \& Astrophysics, ANU,
Canberra, Australia}

\begin{document}

\date{Accepted 2010 October 07. Received 2010 October 06; in original form 2010 September 16}

\pagerange{\pageref{firstpage}--\pageref{lastpage}} \pubyear{2009}

\maketitle

\label{firstpage}
\vspace{-3cm}
\begin{abstract}
As part of our current programme to test $\Lambda$CDM predictions
for dark matter (DM) haloes using extended kinematical
observations of early-type galaxies, we present a dynamical
analysis of the bright elliptical galaxy \N4374\ (M84) based on
$\sim$~450 Planetary Nebulae (PNe) velocities from the
PN.Spectrograph, along with extended long-slit stellar kinematics.

This is the first such analysis of a galaxy from our survey with a radially
constant velocity dispersion profile.
We find that the spatial and kinematical
distributions of the PNe agree with the field stars in the region
of overlap.

The velocity kurtosis is consistent with zero at almost all radii.

We construct a series of Jeans models, fitting both velocity
dispersion and kurtosis to help break the mass-anisotropy
degeneracy. Our mass models include DM halos either with shallow
cores or with central cusps as predicted by cosmological
simulations -- along with the novel introduction in this context
of adiabatic halo contraction from baryon infall.

Both classes of models confirm a very massive dark halo around
NGC~4374, demonstrating that PN kinematics data are well able to
detect such haloes when present. Considering the default
cosmological mass model, we confirm earlier suggestions that
bright galaxies tend to have halo concentrations higher than
$\Lambda$CDM predictions, but this is found to be solved if either
a Salpeter IMF or adiabatic contraction with a Kroupa IMF is
assumed. Thus for the first time a case is found where the PN
dynamics
may well be consistent with a standard dark matter halo. A cored
halo can also fit the data, and prefers a stellar mass consistent
with a Salpeter IMF. The less dramatic dark matter content found
in lower-luminosity ``ordinary'' ellipticals suggests a bimodality
in the halo properties which may be produced by divergent baryonic
effects during their assembly histories.

\end{abstract}


\vspace{-3cm}
\section{Introduction}\label{sec:intro}
The standard cosmological model, the so-called $\Lambda$CDM (cold
dark matter with a cosmological constant; see e.g.
\citealt{2009ApJS..180..225H}), has been challenged by kinematical
measurements of dwarf and spiral galaxies
(\citealt{2005ApJ...634L.145G,2007ApJ...659..149M,2007ApJ...663..948G,2007MNRAS.378...41S,2008MNRAS.383..297S,2008ApJ...676..920K};
but see e.g. \citealt{2009ApJ...697L..38J,2010Natur.463..203G}).
The confrontation of the precictions of the $\Lambda$CDM with
early-type galaxies (ETGs hereafter) is instead more uncertain. On
the one hand, X-rays (see
\citealt{2003ApJ...586..850P,2004MNRAS.354..935O,2006ApJ...646..899H,2009ApJ...706..980J,2010arXiv1007.5322D})
or discrete tracers such as globular clusters (e.g.
\citealt{2009AJ....137.4956R,2010ApJ...711..484S,2010A&A...513A..52S,2010AJ....139.1871W})
confirmed the presence of massive haloes in the most luminous
systems, particularly at the centres of groups and clusters. On
the other hand, ordinary ETGs, probed with planetary nebulae
(PNe), have manifested discrepancies with $\Lambda$CDM
expectations (see e.g. \citealt[hereafter
R+03]{2003Sci...301.1696R}; \citealt[hereafter
N+05]{2005MNRAS.357..691N}) which may be real or due to the
limitations of observations and dynamical analysis.

ETGs are difficult to probe with standard kinematical techniques
(\citealt{2003ApJ...586..850P}; \citealt{2004MNRAS.349..535O};
\citealt{2006MNRAS.370.1797P};
\citealt{2006A&A...448..155B}; \citealt{2007ApJ...667..731P}),
while they are within the reach of the Planetary Nebula
Spectrograph (\PNS; \citealt{2002PASP..114.1234D}) which along
with other instruments is producing large kinematical samples of
PNe in a variety of galaxy types (R+03;
\citealt{2004ApJ...602..685P,2006MNRAS.369..120M};
\citealt[hereafter D+07]{2007ApJ...664..257D};
\citealt{2008MNRAS.384..943N,2008ApJS..175..522M};
\citealt[hereafter N+09]{2009MNRAS.393..329N}; \citealt[hereafter
C+09]{2009MNRAS.394.1249C};
\citealt{2009ApJ...705.1686H,2010ApJ...721..369T}).

One of the main findings emerging from these observations is the
bimodal behavior of ETG
velocity dispersion profiles in the outer regions: steeply falling
and roughly constant (\citealt{2007IAUS..244..289N}; C+09). These
profiles seem to generally (but not perfectly) track the
bimodality of the central regions of ETGs, which fall into the two
classes of disky, fast rotators of ``ordinary'' luminosity, and
boxy, bright slow rotators
\citep{1992MNRAS.259..323C,1996ApJ...464L.119K,2007MNRAS.379..401E}.
The velocity dispersion profiles are shaped by the combination of
orbit structure and mass distribution, but it is still unclear
which of these drives the halo differences between the two galaxy
classes.

In inferring the mass and the orbital structure, the dynamical
modelling of the PN data has been so far focused on intermediate
luminosity systems with declining dispersion profiles (R+03; D+07;
\citealt[hereafter DL+08]{2008MNRAS.385.1729D}; \citealt[hereafter
DL+09]{2009MNRAS.395...76D}; N+09; \citealt{2010arXiv1007.5200R};
cf. \citealt{2009MNRAS.398..561W,2010ApJ...716..370F}). N+09
summarized the results, comparing constraints from PNe with the
ones on group central ``bright'' galaxies from X-rays and globular
clusters, and drew the tentative conclusion that there is a strong
transition between low- and high-concentration DM haloes. Such a
peculiar trend could imply a transition in the role of baryons in
shaping DM haloes, or a problem with the $\Lambda$CDM paradigm
itself (see also N+05, and N+09 for a detailed discussion).

The picture is far from clear and calls for more extensive
analysis. In this paper we investigate the giant galaxy NGC~4374
(M84) using the stellar and PN kinematics data previously
presented in C+09. This is a bright E1 galaxy ($\sim 3 L^*$
luminosity) located in the Virgo cluster core region. It may be
part of a group falling into the Virgo cluster, but it does not
show any signs of being a group-central object. Mass models have
been constructed by \citet[hereafter K+00]{2000A&AS..144...53K},
and \citet[hereafter C+06]{2006MNRAS.366.1126C}
using stellar kinematics within 1~\Re\ (the effective radius enclosing
half the projected light). Extensive ground-based
photometry has been analyzed in \citet{2009ApJS..182..216K}.

NGC~4374 hosts an AGN as demonstrated by X-ray jet emission
(\citealt{2001ApJ...547L.107F}) correlated with two radio lobes
(\citealt{1987MNRAS.228..557L}), and connected to a massive
central black hole (\citealt{1998ApJ...492L.111B}). The hot
interstellar gas in the galaxy is highly disturbed and not
amenable to a standard X-ray based mass analysis
\citep{2008ApJ...686..911F}.

As a representative of the ``bright-ETG'' population with a
flat-dispersion profile (see e.g. C+09), NGC~4374 provides an important
opportunity to investigate the difference between the low-concentrations
inferred from PNe and the high-concentrations from globular clusters
and X-rays.
These tracers have so far been applied to dfferent classes of galaxies,
which suggests the possibility that there are systematic differences in
the mass tracers themselves.
Alternatively, the mass inferences may turn out to be robust to the
type of tracer used, and then should be examined in more detail to see
if they are explainable within the $\Lambda$CDM framework.

\begin{table}
\caption{NGC~4374: basic data for the dynamical analysis.}
\vspace{-0.5cm} \label{tab:galpar}
\begin{center}
\noindent{\smallskip}\\
\begin{tabular}{lll}
\hline
\hline
Parameter & Value  &  Reference \\
\hline
R.A. (J2000)     & 12$^{\rm h}$ 25$^{\rm m}$ 03.7$^{\rm s}$     & NED$^{1}$ \\
Decl. (J2000)     & +12$\degree$ 53$'$ 13$''$     & NED \\
$v_{\rm sys}$           & 1060\kms               & NED\\
$(m-M)_0$   & 31.17\mag               & \citet{2001ApJ...546..681T}$^{2}$\\
$A_V$       & 0.131\mag               &  \citet{1998ApJ...500..525S} \\
$M_V$       & $-22.41\pm0.10 \mag $ & Sec.~\ref{sec:spatdist}\\
$\Re$ from SB fit&$113.5\arcsec\pm 11$      & Sec.~\ref{sec:spatdist}\\
$\Re$ adopted         &$72.5\arcsec\pm 6$      & Sec.~\ref{sec:spatdist}\\
$\sigma_0$    & 284 \kms & HyperLeda$^{3}$\\
\hline
\end{tabular}
\noindent{\smallskip}\\

\begin{minipage}{9.5cm}

NOTES -- (1):
http://nedwww.ipac.caltech.edu/\\
         (2): corrected by -0.16 mag (see \citealt{2003ApJ...583..712J})\\
     (3): http://leda.univ-lyon1.fr/ \citep{2003A&A...412...45P}\\
     (4): adopting our $(m-M)_0$.
\end{minipage}
\noindent{\smallskip}\\
\end{center}
\end{table}

The paper is organized as follows. Section~\ref{sec:sample}
presents the \N4374\ PN system properties like radial density,
velocity dispersion and kurtosis profile, comparing them with the
stellar light surface brightness and kinematical profiles. We
analyze the system's dynamics in Section~\ref{sec:dynamics} and
discuss the results in relation to
previous galaxy analyses in Section~\ref{sec:discuss}. In
Section~\ref{sec:concl} we draw conclusions.
An Appendix covers model variations with an alternative choice of rejected outlier PNe.

\begin{figure}
\hspace{-0.6cm} \epsfig{file=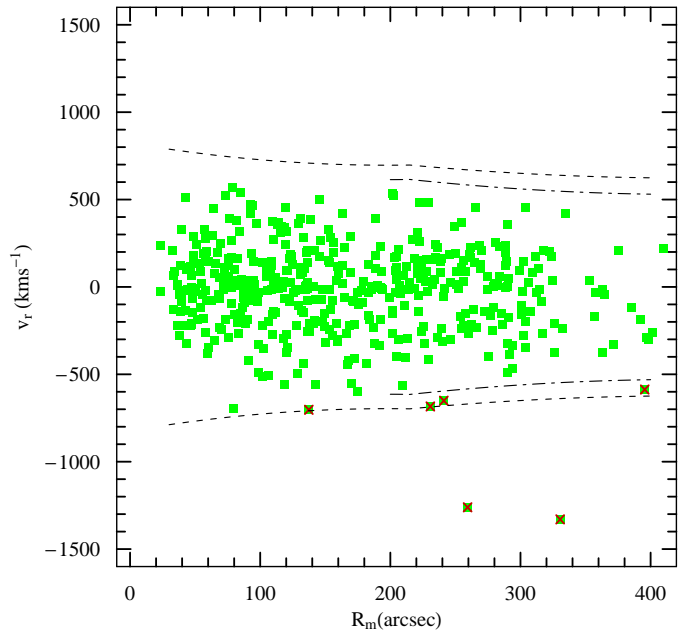,width=8.9cm}
\caption{Distribution of line-of-sight velocities of PN candidates
around NGC~4374, as a function of radius, and relative to the
systemic velocity (1060~\kms). Red $\times$ symbols mark objects
designated as outliers and green boxes show the {\it bona fide}
PNe. The dotted line shows the 3~$\sigma$ velocity envelope. The
dash-dotted line the 3~$\sigma$ velocity envelope corrected by the
energy injected by the interaction with NGC~4406 (see Appendix).}
\label{fig:outli}
\end{figure}

\section{PN system properties}\label{sec:sample}

The data that will be the basis of our dynamical modelling were
presented in C+09, which can be consulted for details of
observations and data reduction. Deep long-slit stellar spectra
were obtained with the VLT+FORS2 spectrograph along the major and
minor axes, and 454 PN candidate velocities with the WHT+PN.S.
Observations were carried out on two different runs (1--4 Apr 2005,
29-3 Mar 2006) with quite uniform seeing conditions ($\sim
1.2''$). To accommodate the anticipated kinematics range for the galaxy, filter
AB at 0\degree\ tilt was used, which has an estimated bandpass of
$\sim5026$\AA\ with $36$\AA\ FWHM.

Here we begin by revisiting some of the data characterization
steps with a few differences optimized for the dynamical analysis.

We present the basic properties of the field stars and PNe in
NGC~4374, including their distributions in space and velocity.
Since an important assumption of our models is that the PNe are a
fair tracer population of the field stars, we compare throughout
the properties of the stars and PNe. The full line-of-sight
velocity field of the PN system of the galaxy has been discussed
in C+09 (see, e.g., their Fig. 3). Both the galaxy light and the
PN distribution appear round so we will assume spherical symmetry
and use, as radial distance from the galaxy centre, the projected
intermediate axis $R_m$, which is related to the semi-major axis
radius $R_a$ and ellipticity $\epsilon$ by $R_m \equiv R_a
(1-\epsilon)^{1/2}$ [where $\epsilon(R_a)$ is taken from
\citealt{2009ApJS..182..216K}].

For the dynamical analysis in this paper, we have concentrated on
identifying possible outliers which could be due to unresolved
background emission-line galaxies, or to PN pair mismatches in
crowded regions or in the case of NGC~4374, to PNe belonging to
the nearby giant elliptical NGC~4406. As done in D+07 and N+09, we
have combined a 3~$\sigma$ clipping criterion plus a
``friendless'' algorithm introduced in
\citet{2003MNRAS.346L..62M}. In Fig.~\ref{fig:outli} we show the
PN individual velocities versus $R_m$ where we have marked with
red crosses the PNe which were either outside the 3~$\sigma$
velocity envelope or turned out to be friendless (i.e. having a
velocity more than $3~\sigma$ away from the average velocity of
their $20$ nearest neighbours).

Using this approach we exclude 6 out of 457 PNe from the C+09
catalog. Some of these 6 differ from the outliers identified by
C+09 because the friendless algorithm is now applied to the raw
data set rather than a point-symmetrized version. The outliers
show a notable asymmetry with respect to the systemic velocity,
which motivates the use of the non-symmetrized friendless
algorithm, and which is probably due to a fly-by encounter of a
nearby giant galaxy as discussed in the Appendix.

The outlier selection is not foolproof
and is a potential source of bias in the analysis.

In the Appendix, we also explore the impact on the dynamical models
of varying the outlier selection,
and find that the mass results are not significantly affected by
changes in the outlier selection, while
the anisotropy inferences are sensitive to the classification of a
small number of objects.
Follow-up spectroscopy of these objects would clearly be valuable.

We next examine the spatial distribution of the final
catalog in \S\ref{sec:spatdist}, and the velocity dispersion and the
kurtosis in \S\ref{sec:dispsec}.

\subsection{Surface photometry and PN spatial distribution}\label{sec:spatdist}

\begin{figure}
\hspace{-0.5cm} \epsfig{file=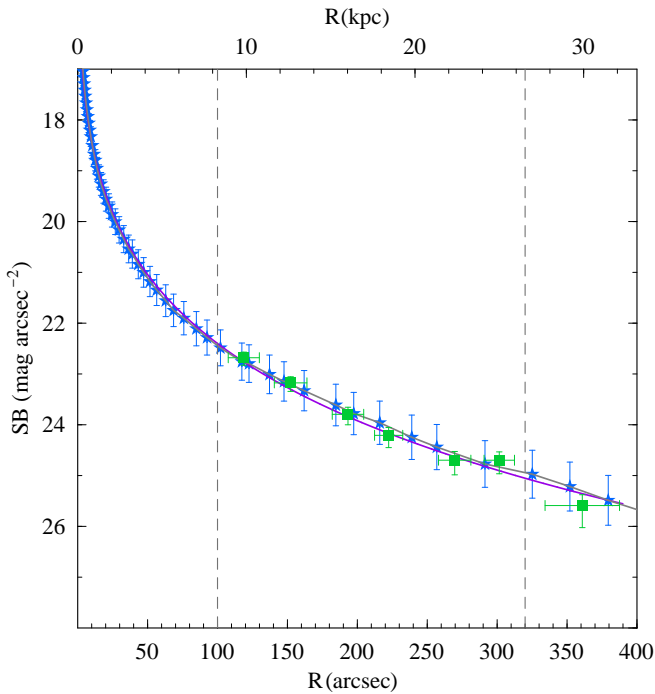,width=8.7cm}
\caption{Radial surface density profiles of the field stars
($V$-band; blue star symbols) and of the PNe (green squares) in
\N4374. The PN number counts have been corrected for spatial
incompleteness, and arbitrarily normalized to match the stellar
data. The vertical error bars of the PN data in this and in the
following figures represent the 1~$\sigma$ uncertainties (based in
this case on counting statistics and completeness correction
uncertainties), while the horizontal error bars show 68\% of the
radial range of the PNe in each bin. The purple curve is a
S\'ersic model fit to the stellar photometry,
and the gray solid curve is the interpolating profile. The vertical dashed lines show the spatial completeness interval of the PN system.}
\label{fig:spatcomp}
\end{figure}

For the galaxy light, we have used the surface photometry from
\citet{2009ApJS..182..216K} as in C+09, but we have reduced the
major/minor axis to a single profile as a function of $R_m$, as
shown in Fig. \ref{fig:spatcomp} [hereafter, $R_m$ and $R$ will be
used interchangeably for the intermediate-axis radius].

To characterize the stellar luminosity profile,
we parametrize the surface brightness (SB)
profile by the S\'ersic law:
\begin{equation}\label{eq:sersic}
\mu(\Rm)-\mu(0) \propto (\Rm/a_S)^{1/m} ,
\end{equation}
where $a_S$ is a scale length and $m$ describes the ``curvature''
of the profile \citep{1968adga.book.....S}. Fitting the dataset
from $R=1''$ to $465''$, we find $a_S=0.00003\arcsec$, $m=6.11$,
and $\mu_0 = 9.74$~mag~arcsec$^{-2}$, and $\Re= 113.5\arcsec$.
Other values obtained in the literature include $51''$ from
\citet{1991trcb.book.....D},
$53''$ from \citet{2001MNRAS.327.1004B},
$72.5''$ from C+06,
$93''$ from \citet{2009ApJS..181..486H},
$114''$ and $142''$ from (\citealt{2009ApJS..182..216K}; along the semi-major axis),
and $204''$ from \citet{2010ApJ...715..972J}.

These differences do not mean that the galaxy's luminosity profile
is not reasonably well known over the region which we will be
modelling, but that for high-S\'ersic-index galaxies, certain
characteristic quantities such as \Re\ and total luminosity
require considerable extrapolation and are poorly constrained.

This is not a problem that we will solve overnight, and for the sake
of using an \Re\ parameter that is equivalent to the most common usage in
observations and theory, we adopt $\Re=72.5''$ from the wide-field $R^{1/4}$
growth-curve fitting of C+06.
This differs from our approach in C+09, where for the sake of uniformity
we adopted the \citet{2001MNRAS.327.1004B} values, which will not be reliable
for very extended galaxies because of the narrow imaging fields used.
Our modelling will all be conducted in physical units, so this choice
of \Re\ impacts only the quoting of radial ranges in some cases, and
the comparisons to simulated galaxies.

With our full S\'ersic solution, the extinction-corrected total
luminosity in the $V$-band is $7.64\times10^{10} L_{V, \odot}$, or
$M_V = -22.4$; the uncertainties in the outer surface brightness
profile yield a (model-dependent) total luminosity uncertainty of
$\sim$10--15\%.
These and other global parameters for \N4374\ are listed in Table
\ref{tab:galpar}.
%
For practical use in the Jeans modeling, we have also produced a smoothed
density profile from the data
made by a combination of a simple interpolation of the data up to $290''$
and our S\'ersic model outside this radius.

We next compare the spatial density of the PNe with the field
stars, using the PN number density complete to $m^*+1.1$ (see
C+09, Table 7). Note that while C+09 used $R_a$, we bin the data using $R_m$.

Given an arbitrary normalization, the {\it PN profile matches the
stellar photometry remarkably well} (Fig.~\ref{fig:spatcomp})---as
also generally found in a larger sample of galaxies by C+09.

\subsection{The dispersion and kurtosis profiles}\label{sec:dispsec}
The rotation and velocity dispersion along the major and minor
axes of \N4374\ have been discussed in C+09 (their fig. 7),
together with the 2D radial velocity field (their fig. 3). For the
spherical analysis in this paper, we reduce these data to a single
average velocity dispersion profile after having re-scaled the two
axes to the intermediate-axis radius $R_m$.

To obtain the azimuthally averaged profile, the rotation and true
dispersion profile are folded into an root-mean-square velocity
profile $v_{\rm RMS}=\sqrt{v^2 + \sigma^2}$\footnote{In the
following we will use spherical Jeans equations for non-rotating
systems. Although NGC~4374 has no significant rotation, the use of
the $v_{\rm RMS}$ will ensure that there is no rotation
contribution missing in the equilibrium balance.}, where $v$ and
$\sigma$ are the rotation and dispersion components
respectively\footnote{In the long-slit stellar data, $v$ and
$\sigma$ are not the true classical moments but fit parameters in
a Gauss-Hermite series which includes the higher-order moments
$h_3$ and $h_4$. In principle, we should convert these fit
parameters into revised estimates of the classical moments, e.g.
using equation 18 of \citet{1993ApJ...407..525V}. Doing so would
lower the outer stellar dispersion profiles by $\sim$~10\%.
However, it is notoriously difficult to extract reliable
measurements of higher-order moments (e.g.
\citealt{2006MNRAS.370..559S}), and we are not confident that the
$h_4$ measurements in this case are accurate.
To avoid introducing spurious corrections to the kinematics, we
therefore assume the $v$ and $\sigma$ fit parameters are good
estimates of the classical moments.}.
This RMS velocity is a measure of the total kinetic energy, and we
henceforth loosely refer to it as the velocity dispersion or
\VD{}. We combine the stellar data from the different axes by
averaging, while folding the (small) systematic differences into
the final uncertainties\footnote{The uncertainties in the PN
dispersion use a classical analytic formula that assumes a
Gaussian distribution, i.e. $\Delta v_{\rm RMS} \sim
\sqrt{\Sigma_i v_i^{2}/2N^2}$. We expect this approximation to
produce accurate results in realistic systems
\citep{2001A&A...377..784N}, and we have carried out additional
Monte Carlo simulations of a simplified galaxy with radial orbits,
finding that the dispersion is very accurately recovered with our
estimator, with a possible bias to be $\sim5\%$ too high.}. The PN
VD is calculated using a classical expression for the variance of
the discrete velocities around the systemic velocity. Note that
the rotation amplitude of $\sim$~50~\kms\ is not dynamically
significant compared to the dispersion of $\sim$200--250~\kms.

\begin{figure}
\hspace{-0.18cm}
\epsfig{file=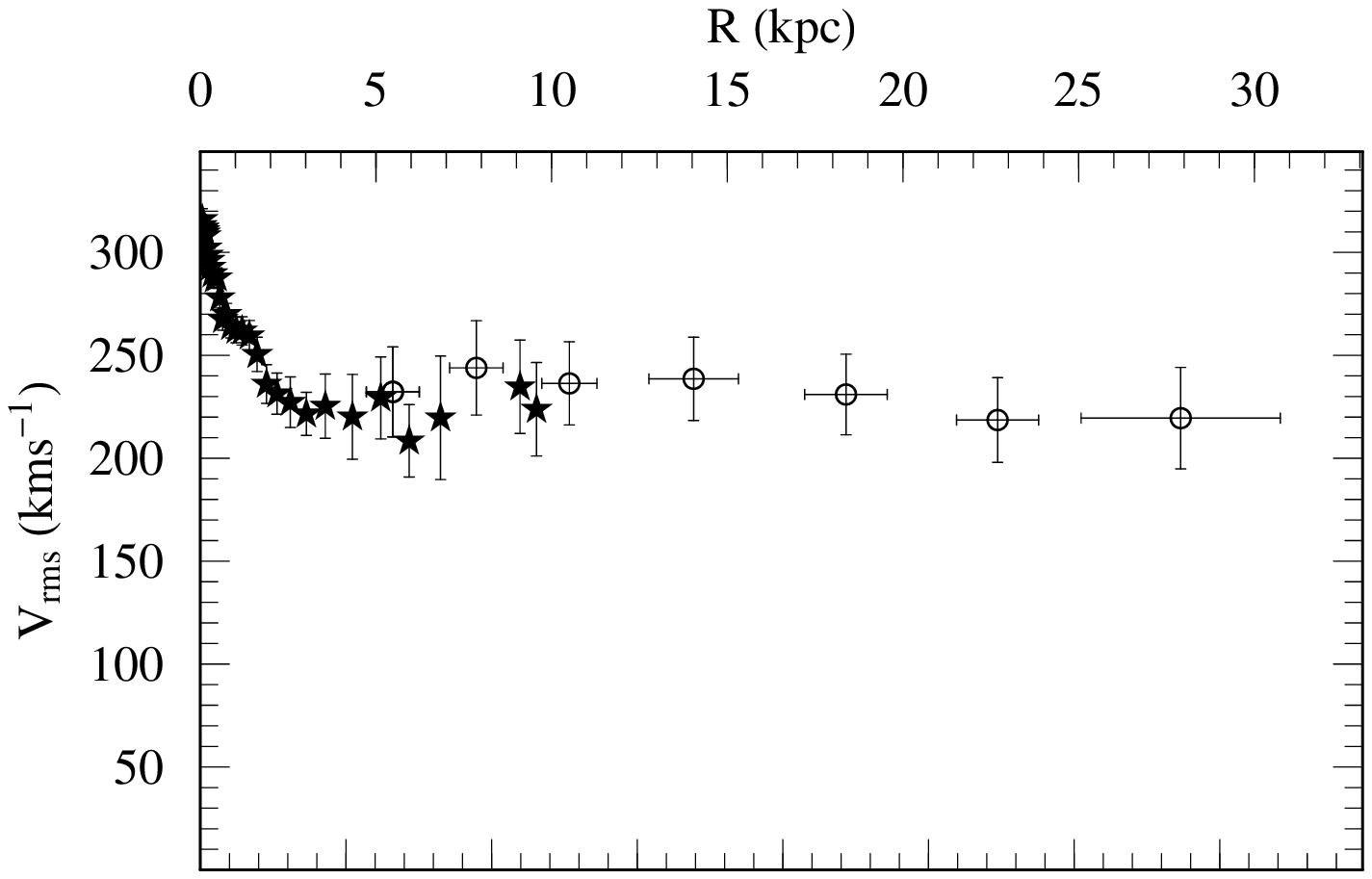,width=8.5cm}

\vspace{0.2cm} \hspace{-0.09cm}
\epsfig{file=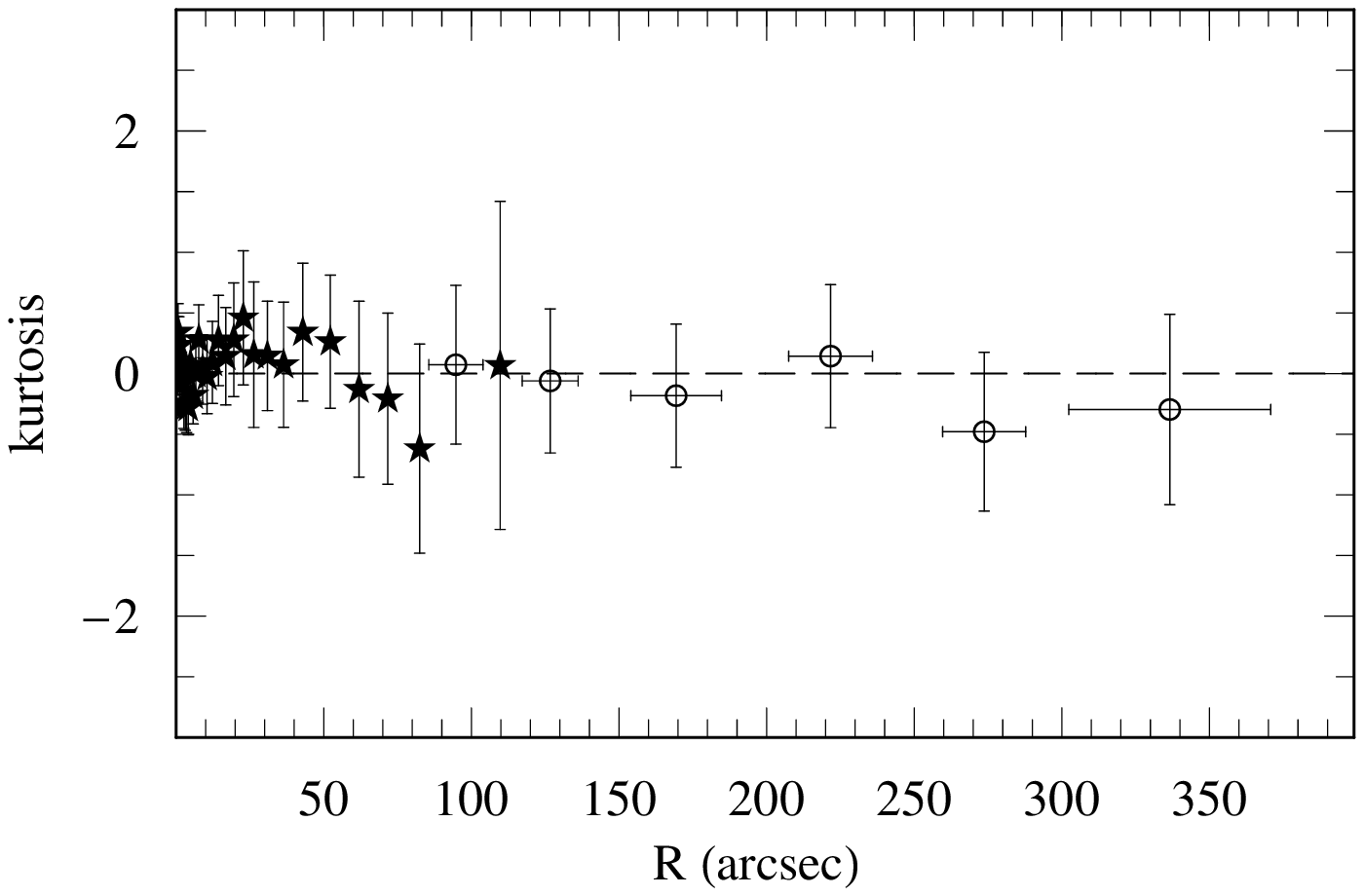,width=8.49cm}
\caption{Composite projected RMS velocity and kurtosis profiles of
NGC~4374, with data from stars (filled star symbols) and PNe (open
circles). Separated profiles of rotation and true dispersion can
be seen in C+09.} \label{fig:dispprof}
\end{figure}

The resulting ``dispersion'' data are plotted in
Fig.~\ref{fig:dispprof}. Overall, the use of the full PN sample in the
azimuthal averaged profile allows us to map the kinematics of the galaxy
out to $\sim340''$ which is 20\% farther out than the major/minor axis
analysis performed in C+09.

The dispersion decreases sharply from the centre out to 50$''$
where the \VD\ from the long slit data flattens at $\sim 220
\kms$. {\it The PN data are consistent with the stellar absorption
estimates in the region of overlap} and possibly show a rise of
the \VD\ profile from 100$''$ with a peak of $\sim240$ \kms at
170$''$ (corresponding to about 15 kpc for our adopted distance)
and a subsequent decrease to the original value of 220 \kms\ where
the \VD\ stays flat out to the last data point ($\sim340''$ or 27
kpc, i.e. $\sim$5\Re). This makes NGC 4374 a prototypical system
with a flat dispersion profile, although the uncertainty on the
\Re\ estimate
(e.g. for $\Re=204''$ the last data point is at $\sim$1.8\Re)
provides a warning that we may not be sampling far enough from the
center to probe the full dynamical range of the system. More
extended data (ideally in the direction opposite to the nearby
galaxy NGC 4406 where it is more likely that the stellar
kinematics is undisturbed) would clarify whether the velocity
dispersion remains flat farther out, starts to decrease as
observed in the intermediate luminosity sample or begins to rise
as the cluster potential is probed.

The $v_{\rm RMS}$ shows a bump at around 15$''$ which is also seen
in the major and minor profiles and might be related to some
kinematical substructure\footnote{This may be related to the
central dust ring clearly seen in optical imaging of the galaxy
\citep{1994AJ....108.1567J}.} which is not evident in the
photometric profile (see Fig. \ref{fig:spatcomp}). As we are
mainly interested in modelling the galaxy outskirts, the presence
of these wiggles in the kinematical profile will not affect our
analysis.

The nearly flat dispersion profile in Fig. \ref{fig:dispprof}
corresponds to an asymptotic slope of $-0.07\pm0.07$ which is in
clear contrast with the decreasing profiles found in intermediate
luminosity galaxies, with typical power-law exponents of
$-0.2$ to $-0.6$ (see R+03; D+07; N+09).
\citealt{2007IAUS..244..289N} and C+09 identified a possible
dichotomy of early-type galaxies based on these dispersion slope
differences, and we will here investigate further the dynamical
implications for \N4374.

We next consider higher-order velocity information.
We quantify the
shapes of the stellar and PN line-of-sight velocity distirbutions
(LOSVDs) in \N4374\ using a
classical dimensionless kurtosis,
$\kappa\equiv\overline{v^4}/(\overline{v^2})^2-3$
(see \citealt{Joanes98} for exact expressions and
uncertainties\footnote{Monte Carlo simulations based on
\citet{2001A&A...377..784N} models have demonstrated accurate
recovery of the kurtosis using our estimator, with a systematic
deviation of no more than $\sim 0.1$, see also N+09.}). Broadly
speaking, we can expect that $\kappa \simeq 0$ is a fair
indication of isotropic orbits, $\kappa < 0$ is pertinent to
tangential orbits and $\kappa > 0$ for radial orbits.

In Fig.~\ref{fig:dispprof} we have combined the PN estimates with
the stellar equivalent by converting the long slit stellar
Gauss-Hermite coefficient $h_4$
\citep{1993ApJ...407..525V,1993MNRAS.265..213G} into kurtosis
estimates using the approximate relation $\kappa\simeq 8\sqrt{6}
h_4$.

The PN kurtosis is consistent with the stellar properties
in the region of overlap. Thanks to the large statistical sample,
the PN data points show error bars which are fairly similar to stellar
estimates, based on the best quality stellar absorption line data.
The total kurtosis profile is consistent with zero at all radii
and has a median (calculated over all datapoints) of $0.05\pm0.19$.

Our previous analyses of NGC~3379 and NGC~4494 indicated global
$\kappa \sim +0.2$ and $+0.6$, respectively. However, most of this
difference is driven by the data inside \Re, where previous work
with larger galaxy samples has indicated that any correlations
between the fourth moment and other galaxy properties are subtle
\citep{1994MNRAS.269..785B,2008MNRAS.390...93K}.

In the outer parts, all three galaxies are similarly consistent
with zero kurtosis, and it will be interesting to see if any
patterns emerge with a large sample. However, as we will see in
the next Section, interpreting the orbital anisotropy implications
of the kurtosis requires detailed modelling.

\section{Dynamical Models}\label{sec:dynamics}

We present a suite of Jeans dynamical models following the same
scheme as in N+09, to which we refer the reader for more details
of the analysis. We will combine the photometric and kinematical
data for the stars and PNe in \N4374\ into integrated models in
order to derive the mass profile and the orbital distribution of
the galaxy and finally test whether or not it hosts a massive dark
halo compatible with the \LCDM\ predictions.

Although there are other dynamical procedures such as
Schwarzschild's method and made-to-measure particle methods (e.g.
R+03; \citealt{2008ApJ...682..841C}; DL+08) that have been applied
to discrete velocity data and are more robust than our Jeans
approach, the latter is computationally faster and somewhat more
intuitive. Furthermore it allows a much larger flexibility on the
range of galaxy potentials to be used. In the following we briefly
remind the main steps of our dynamical procedures.

In the different formulations of the Jeans equations we will
assume spherical symmetry. This is a reasonable approximation
because the round and boxy stellar isophotes of \N4374\ (average
ellipticity $\langle\epsilon \rangle=0.13$ and $\langle a_4
\rangle=-0.4$; see C+09), and the small $V/\sigma=0.03$
(\citealt{2007MNRAS.379..418C})\footnote{Their fig.~3 illustrates
an estimated family of deprojections for this galaxy, with the
most flattened solution having $\epsilon \sim 0.2$.} make the
system a typical boxy--slow rotator which is highly unlikely to be
very flattened intrinsically.

Another basic assumption of our analysis is that stars and PNe are
all the drawn from the same underlying dynamical tracer
population, which is well motivated by the agreement between the
stellar and PN properties (\S\ref{sec:spatdist} and
\ref{sec:dispsec}). We will also in general omit the stellar
kinematics data inside 10$''$ from our model fits, since there
appears to be a strong dynamical change in the nuclear region
which our smooth Jeans models are not designed to reproduce
(partially produced by a massive black hole;
\citealt{1998ApJ...492L.111B}\footnote{Here
\citet{1998ApJ...492L.111B} estimate a black hole mass of $M_{\rm
BH} \sim 1.5 \times 10^9 \Msun $ which implies a sphere of
influence of radius $r_h \sim 1.7''$, where we have defined $r_h$
as the radius where $M_*(r < r_h ) = 2M_ {\rm BH}$, with $M_*(r)$
corresponding to the Kroupa IMF.}).
\begin{figure}
\hspace{-0.9cm} \epsfig{file=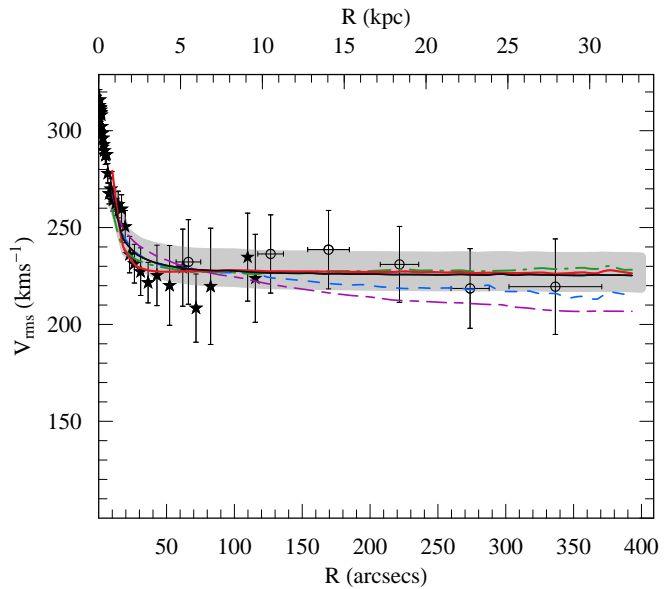,width=9.8cm}
\caption{Composite projected velocity dispersion profile of
NGC~4374, with data from stars (filled star symbols) and PNe (open
circles). The black solid curve shows the pseudo-inversion mass
model to fit the PN data outside 10$''$ for the isotropic case,
with the shaded regions showing the 1 $\sigma$ significance of the
fit. The short dashed blue curve shows the solution for
$\beta=0.5$, the dot-dashed green curve the one for $\beta=-0.5$.
The long-short-dashed violet line shows the solution for the
cosmological motivated $\beta(r)$ profile as in Eq.
\ref{eq:MLbeta}. The thick red solid line shows the heuristic
$\beta(r)$ model adopted in Sect. \ref{sec:resNFW}.}
\label{fig:dispprof2}
\end{figure}

We begin with a simple non-parametric model in
\S\ref{sec:massinv}, then introduce multi-component mass-models in
\S\ref{sec:massmod} and additional dynamical methods in
\S\ref{sec:dynmeth}. The multi-component results are presented in
\S\ref{sec:dynres}--\ref{sec:resLOG} and the mass profiles
summarized in \S\ref{sec:mass_vcirc}.

\subsection{Pseudo-inversion mass model}\label{sec:massinv}
We start with a phenomenological approach introduced in R+03 and
followed in D+07 and N+09, used to convert the observed kinematics
into a mass profile $M(r)$.  This approach has the advantage that
it is computationally light, does not involve Abel inversion
integrals, and does not assume any form for $M(r)$, nor a stellar
$M/L$ value (which will be discussed later in this Section). A
disadvantage is that it does not allow a direct test of any
theoretical prediction (which we will do in the next Sections).

For the benefit of readers not familiar with this procedure, we
summarize in the following its basic steps:
\begin{enumerate}
\item Adopt a simple smooth parametric function for the intrinsic radial velocity dispersion profile:
\begin{equation}
\sigma_r(r) = \sigma_0 \left[1+\left(\frac{r}{r+r_0}\right)^\eta \right]^{-1},
\label{eq:v0eqn}
\end{equation}
where ${\sigma_0,r_0,\eta}$ are a minimalistic set of free
parameters. This model is adopted to reproduce the flat dispersion
profile in the outer galaxy regions and is different from those
adopted in D+07 and N+09 which were constructed to match steeply
decreasing velocity dispersion profiles.
\item Assume a given anisotropy profile, often constant or parametrized
as a simple function:
\begin{equation}\label{eq:beta}
\beta(r) \equiv 1-\sigma^2_{\theta}/\sigma^2_r ,
\end{equation}
where $\sigma_\theta$ and $\sigma_r$ are the spherically-symmetric
tangential and radial components of the velocity dispersion
ellipsoid, expressed in spherical coordinates\footnote{Due to the
modest rotation of the galaxy, we expect the spherical
approximation not to cause any significant systematic issues.}.
\item Project the line-of-sight components of the
3-D velocity dispersions $\sigma_r$ and $\sigma_\theta$
for comparison with the line-of-sight velocity dispersion data $\sigma_{\rm los}(R)$.
\item Iteratively adjust the free parameters in Eq.~\ref{eq:v0eqn},
to best fit the model to the observed dispersion profile.
\item Use the best-fit model (Eq.~\ref{eq:v0eqn}) in the Jeans equation 4-55 of
\citet{1987gady.book.....B} to calculate $M(r)$:
\begin{equation}
M(r) = -\frac{\sigma_r^2~r}{G}\left( \frac{d\ln j_*}{d\ln r}+\frac{d\ln\sigma_r^2}{d\ln r} +2\beta \right) ,
\label{eq:masseq}
\end{equation}
where $j_*(r)$ is the spatial density of the PNe, and corresponds
to an Abel deprojection of a smoothed density law as in \S\ref{sec:spatdist}.
Additional quantities may then be computed, such as the cumulative $M/L$.
\end{enumerate}

\begin{figure}
\hspace{-0.7cm} \epsfig{file=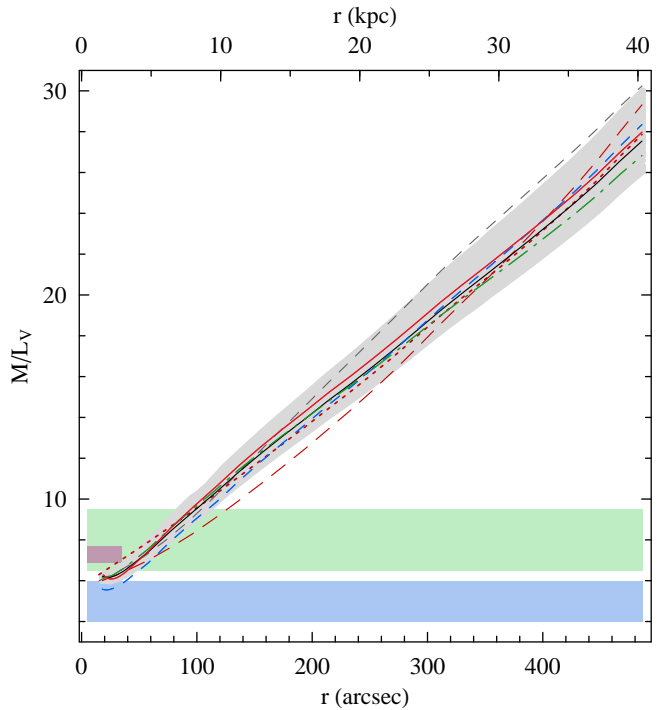,width=9.5cm}
\caption{Cumulative $V$-band mass-to-light ratio ($M/L$) of
NGC~4374 (note that the vertical axis starts from $\Ysol=3$). The
curves based on the pseudo-mass inversion method are colour coded
as in Fig. \ref{fig:dispprof2}. We also add some of the models
from the Jeans analysis in \S\ref{sec:dynmeth}: the dotted red
curve is the ``NFW + $\beta(r)$'', the dashed red curve is the
same model with adiabatic contraction [``NFW + AC + $\beta(r)$''],
and the dashed gray line is the logarithmic potential model with
$\beta(r)$ (see \S\ref{sec:resNFW} and \S\ref{sec:resLOG}). The
horizontal blue shaded region shows the stellar $M/L$ and its
uncertainty for the Kroupa IMF, while the green one is for the
Salpeter IMF. The small purple shaded region is the dynamical
$M/L$ estimate from C+06. See text for details.}
\label{fig:modelfitstwo}
\end{figure}

Starting with the isotropic case ($\beta=0$), we find that the
simple model~(\ref{eq:v0eqn}) is able to fit the dispersion data
well (Fig.~\ref{fig:dispprof2}), with some systematic discrepancies
at $\sim$~$40''$ that we will improve upon with more complicated models below.
The resulting $M/L$ profile
increases steeply with the radius (Fig.~\ref{fig:modelfitstwo}),
providing a strong indication for the presence of an extended DM
halo.
Note that the shaded regions in Figs.~\ref{fig:dispprof2} and \ref{fig:modelfitstwo}
along with the various uncertainties quoted below account for the
1-$\sigma$ statistical confidence region in the parameter space
(${\sigma_0,r_0,\eta}$) of the dynamical model.

The central dynamical $(\ML)_V=6.5$ can also be compared with
independent stellar population analyses of the
stellar $M/L$, $\Upsilon_*$.
Assuming a \citet{2001MNRAS.322..231K} IMF,
\citet{Tortora+09} found
$\Upsilon_*\sim$~3--4.5~$\Ysol$,
while \citet{2001AJ....121.1936G} found
$\Upsilon_*\sim$~4.5--6.0~$\Ysol$
(where we have in both cases converted from $B$- to $V$-band).
C+06 found
$\Upsilon_*=3.08\Upsilon_\odot$ in $I$-band which
we convert to
$\Upsilon_*\simeq$~5.14~$\Ysol$ after detailed comparison of the SB profiles.
(Note that their Schwarzschild modelling analysis implies a dynamical
$\Upsilon \simeq 7.3 \pm 0.4 \Ysol$ in the central regions, which agrees with
our Jeans results, as shown in Fig.~\ref{fig:modelfitstwo}.)

We can reasonably assume $\Upsilon_*\sim$~4--6~$\Ysol$ for a
Kroupa IMF, which corresponds to $\sim$~6.5--9.5~$\Ysol$ for a
\citet{1955ApJ...121..161S} IMF (see Fig. \ref{fig:dispprof2}).
Therefore the dynamical $M/L$ is suggestive of some dark matter inside \Re\
($72.5''$)
for the case of Kroupa but not Salpeter.
In the following we will consider the stellar $M/L$ based on the
Kroupa IMF as the reference results, since there are arguments to
consider this one as a universal IMF (Kroupa 2001).

Our last datapoint ($\sim340''$) is close to $\sim 5\Re{}$,
which is a benchmark distance for the mass profiles (see R+03,
D+07 and N+09): here we find that the $V$-band $M/L$ within this
radius is $\Upsilon_{{\rm 5}, V} \sim 20\pm2$~
$\Upsilon_{\odot,V}$\footnote{Hereafter we are deliberately
neglecting the uncertainty on \Re\ which we have seen are
unreasonably large and scale all the results for our assumed
\Re.}. The anisotropy is accounted for in step (ii) of the
procedure by adopting constant values of $\beta=\pm$~0.5 as a
plausible (though not exhaustive) range of the stellar anisotropy.
The fits to the data are just as good as for the isotropic case as
shown in Fig.~\ref{fig:dispprof2}.

In Fig.~\ref{fig:modelfitstwo} we show the $M/L(r)$ profiles
corresponding to the three $\beta$ values. Assuming $\beta=+0.5$
implies a smaller central $\ML$ ($\sim$~5.5~$\Ysol$) but a steeper
$\ML$ profile than the case of $\beta=0$, while $\beta=-0.5$
implies a larger central $\ML$ ($\sim7\Ysol$) and a shallower
$\ML$ profile outside $1\Re$ (with the $M/L$ consistent with the
isotropic profile at all radii in either cases). In all cases a
constant $M/L$ is excluded at more than 3~$\sigma$ and DM starts
dominating already at $1\Re$, assuming a Kroupa IMF, and at $\sim
2\Re$ for the Salpeter IMF case.

Our outer $\ML$ results are relatively insensitive to the anisotropy assumed
because of a geometrical effect in certain regimes in radius
that causes anisotropy differences
to cancel out when projected to line-of-sight velocity dispersions
(cf. \citealt{1993MNRAS.265..213G}, Fig.~8;
\citealt{1994MNRAS.270..271V}, Figs. 10 and 11; \citealt{2010MNRAS.406.1220W}).
This ``pinch point''
 occurs where the 3D log slopes of the tracer density profile $\alpha$ and the
velocity dispersion $\gamma$ add up to $(\alpha+\gamma) \simeq -3$
(see \citealt{2005Natur.437..707D} Eq.~2). In a bright elliptical
galaxy like N4374, the high Sersic index $n$, the large
scale-length, and the flat dispersion profile combine to push the
pinch point to fairly large radii: $\sim$~100$''$ in this case.
This robustness of the mass inference contrasts with the case of
galaxies with steeply declining dispersion profiles, where the
mass-anisotropy degeneracy is particularly severe (DL+08; DL+09).

\begin{figure}
\hspace{-0.7cm} \epsfig{file=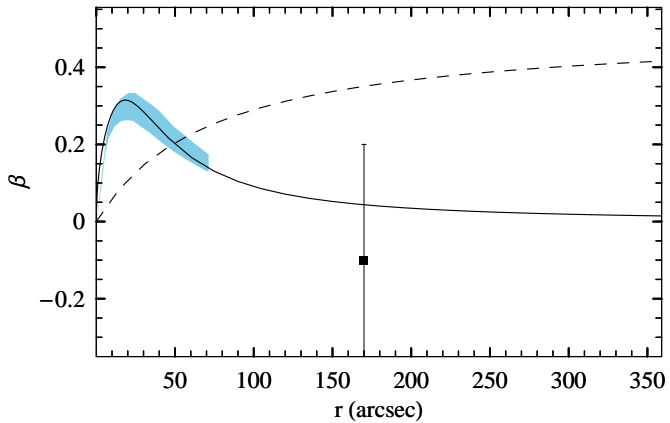,width=9.cm}
\caption{The heuristic $\beta(r)$ profile in Eq. \ref{eq:betar2}
(solid line) is compared with the simulation based $\beta(r)$
(dashed line) from M{\L}05 as in Eq. \ref{eq:MLbeta} and the
modelled $\beta(r)$ from \citealt{2000A&AS..144...53K} (shaded
region). Here only the radial range covered by the Kronawitter et
al. model is shown: the matching with the heuristic $\beta(r)$ is
good, while the M{\L}05 formula predicts radial anisotropy at much
larger distances from the centre. The anisotropy value derived
from direct kurtosis inferences (see \S\ref{sec:dynmeth}) is also
shown with 1 $\sigma$ error bars.} \label{fig:beta_vs_kron}
\end{figure}

We have also tested the anisotropy profile based on theoretical
expectations from merging collisionless systems as derived from
M{\L}05:
\begin{equation}\label{eq:MLbeta}
\beta(r)= \beta_0 {r\over r+r_{\rm a}} ,
\end{equation}
where $\beta_0 \simeq 0.5$ and $r_{\rm a}\simeq1.4\Re$ (based on
the merger simulations of D+05). Adopting this profile with
$r_{\rm a}=101''$, we find that the \VD\ profile matches slightly
better the central regions, but it fits poorly the large radii
datapoints. In this respect
Eq.~\ref{eq:MLbeta} seems to be ineffective in reproducing the
intrinsic anisotropy of the galaxy (given the limits of the simple
parametrization assumed in equation~\ref{eq:v0eqn})\footnote{We
tried out a wider
range of $r_{\rm a}$: for smaller $r_{\rm a}$ the predicted
dispersion was still lower than the data, and for
larger $r_{\rm a}$ the dispersion progressively approached the
isotropic case.}
However, the fact that radial anisotropy produces a better fit to
the central VD while $\beta=0$ matches the outer parts of the
galaxy suggests that a more complicated $\beta(r)$ profile than
the one in Eq. \ref{eq:MLbeta}
should be applied to \N4374.\\
For instance, looking at the kurtosis profile in Fig.
\ref{fig:dispprof}, one suspects that a $\beta(r)$ profile
which is isotropic in the very central regions ($R<5''$) and in
the outer parts ($R\gsim70''$) and radially anisotropic in
between ($R\sim 5''-70''$) might do a better job. Following this
heuristic approach we adopt the following formula:
\begin{equation}\label{eq:betar2}
\beta(\xi)= \beta_0 \frac{\xi^{1/2}}{\xi^2+1} ,
\end{equation}
where $\beta_0 =0.6$ and $\xi=r/r_{\rm a}$ with $r_{\rm
a}\sim30''$ (see also Section~\ref{sec:resNFW}). This $\beta(r)$
profile is significantly different from the simulation-based Eq.
\ref{eq:MLbeta} but similar to the $\beta(r)$ found from the
detailed dynamical models of K+00 for \N4374\ (see Fig.
\ref{fig:beta_vs_kron}) as well as for some other galaxies in
their sample (e.g.
NGC~4278, NGC~4472, NGC~4486, NGC~5846).
In this case the best fit to the VD is improved as shown in
Fig. \ref{fig:dispprof2} (red curve; we will come back to this issue in
\S\ref{sec:resNFW}). The corresponding $M/L$ profile has a
central value which is closer to the isotropic solution
($\sim6.5\Ysol$) and becomes slightly larger outward, finally
converging to the isotropic case asymptotically.

The overall plausible range for the benchmark-radius $M/L$ of
NGC~4374 is $\Upsilon_{{\rm 5},V}=$~18--24~$\Upsilon_{\odot,V}$
(including both statistical uncertainties as well as the
systematic anisotropy uncertainties, given the range of $\beta(r)$
profiles that we allow). This result is significantly larger than
the typical $\ML$ found for the intermediate-luminosity galaxy
sample studied so far with the PN.S (see e.g. R+03, D+07, DL+09
and N+09), but more similar to the $\ML$ estimates found in bright
systems using globular clusters and X-rays (e.g.
\citealt{2006ApJ...646..899H,2009AJ....137.4956R,2010A&A...513A..52S,2010arXiv1007.5322D}).

The steep increase of the $M/L$  with radius can be
quantified through the dimensionless $\ML$ gradient
(introduced by N+05):
\begin{equation}
\dML \equiv \frac{\Re \Delta \Upsilon}{\Upsilon_{\rm in} \Delta R} ,
\end{equation}
where $\Upsilon_{\rm in}$ is the central dynamical $M/L$. For
\N4374\ we find $\dML =$~0.5--0.7, which places this galaxy among
the systems with larger \dML\ which are discussed in N+05 as very
dark-matter dominated.
As a comparison, for NGC~3379 and NGC~4494 we found
 $\dML$ in the range $-0.05$ to 0.25.

\subsection{Multi-component models: mass profiles}\label{sec:massmod}

The second strategy for our dynamical analysis again uses a Jeans
analysis but begins with parameterized mass profiles and projects
the predicted kinematics for comparison to the data. Following
N+09, the inclusion of higher velocity moments (kurtosis) in the
Jeans analysis is expected to alleviate the mass-anisotropy
degeneracy.

In our equations, we will adopt two-component mass models
consisting of a luminous field star distribution plus a DM halo.
%
The total gravitational potential may thus be expressed as
$\Phi=\Phi_*+\Phi_{\rm d}$. The stellar gravitational potential
$\Phi_*(r)$ is derived from the stellar luminosity
$j_*(r)$\footnote{This is obtained by Abell inversion of the
observed SB in the central regions and the extrapolation to
infinity according to the S\'ersic model of
\S\ref{sec:spatdist}.}, combined with some assumed constant
$\Upsilon_*$.

Our mass models as described below use for the DM either an
NFW profile (\S\ref{sec:lcdm}) or a pseudo-isothermal form (\S\ref{sec:LOG}).

\subsubsection{NFW model}\label{sec:lcdm}

Our reference mass models aims at testing the predictions from
simulations of collisionless DM halo formation in a $\Lambda$CDM
cosmology. In this case the DM density takes the approximate form
of an NFW profile:
\begin{equation}
\rho_{\rm d}(r)=\frac{\rho_s}{(r/r_s)(1+r/r_s)^2} ,
\label{rhoNFW}
\end{equation}
where $\rho_s$ and $r_s$ are the characteristic density and scale radius of the halo.
The cumulative dark halo mass is
\begin{equation}
M_{\rm d}(r)=4 \pi \rho_s r_s^3 A(r/r_s) ,
\end{equation}
where
\begin{equation}
A(x) \equiv \ln (1+x)-\frac{x}{1+x}.
\end{equation}
The potential is:
\begin{equation}
\Phi_{\rm d}(r) = \frac{4\pi G \rho_s r_s^3}{r} \ln \left(\frac{r_s}{r+r_s}\right) ,
\end{equation}
where $G$ is the gravitational constant.

The three free parameters describing the NFW mass model are thus
$\Upsilon_*$, $\rho_s$ and $r_s$. The halo can alternatively be
parametrized by the virial mass and concentration, $M_{\rm
vir}\equiv 4\pi\Delta_{\rm vir}\rho_{\rm crit}r_{\rm vir}^3/3$ and
$c_{\rm vir}\equiv r_{\rm vir}/r_s$, where the critical density is
$\rho_{\rm crit}=1.37\times10^{-7} M_\odot$~pc$^{-3}$ and the
virial overdensity value is $\Delta_{\rm vir}\simeq 100$.

The expected values for these model parameters are not arbitrary
in $\Lambda$CDM. For instance, in a collisionless $\Lambda$CDM
universe with WMAP5 parameters, the following mean relation is
expected between mass and concentration\footnote{For sake of
completeness we also report here the WMAP1 equations (see N+09 for
details):\\
$
c\vir(M\vir)\simeq18\left(\frac{M\vir}{h^{-1}10^{11}\Msun}\right)^{-0.125}
$ \\
and\\
$ \rho_s \simeq \left(\frac{r_s}{10 {\rm pc}}\right)^{-2/3} \Msun
{\rm pc}^{-3} . $}:
\begin{equation}
c\vir(M\vir)\simeq12\left(\frac{M\vir}{10^{11}\Msun}\right)^{-0.094}
..
\label{cMvir2}
\end{equation}
which has a 1~$\sigma$ scatter of  0.11 dex, and is valid for
$z=0$, $\Omega_m=0.3$, $\Omega_{\Lambda}=0.7$, $h=0.7$,
and $\sigma_8=0.8$
\citep{2008arXiv0805.1926M}.
For comparing with models parameterized by the scale radius $r_s$
and density $\rho_s$ (e.g. Eq.~\ref{rhoNFW}), we find that
Eq.~\ref{cMvir2} is equivalent to the following relation:
\begin{equation}
\rho_s \simeq 0.29 \left(\frac{r_s}{10 {\rm pc}}\right)^{-0.53}
\Msun {\rm pc}^{-3} .
\label{eq:rho_rs_w5}
\end{equation}
where the scatter in $\rho_s$ at fixed $r_s$ is a factor of 1.3.
Note that in N+09 we used $\Lambda$CDM halo predictions based on
WMAP1 parameters, which implied $\sim$~30\% higher concentrations
than WMAP5.

\subsubsection{LOG model}\label{sec:LOG}

Our alternative mass model consists of a logarithmic potential
(\citealt{1987gady.book.....B} \S2.2.2) which was motivated by
observations of spiral galaxy rotation curves (see e.g.
\citealt{1996MNRAS.281...27P}). The potential is:
\begin{equation}
\Phi_{\rm d}(r)=\frac{v_0^2}{2}\ln(r_0^2+r^2) ,
\label{potLOG}
\end{equation}
where $v_0$ and $r_0$ are the asymptotic circular velocity and
core radius of the halo. The corresponding DM density and
cumulative mass profiles are respectively:
\begin{equation}
\rho_{\rm d}(r)=\frac{ v_0^2 (3 r_0^2+r^2)}{4\pi G (r_0^2+r^2)^2} ,
\label{rhoLOG}
\end{equation}
and
\begin{equation}
M_{\rm d}(r)=\frac{1}{G}\frac{v_0^2 r^3}{r_0^2+r^2} .
\label{massLOG}
\end{equation}
The three free parameters of this ``LOG'' model are thus
$\Upsilon_*$, $v_0$, and $r_0$. We define a virial mass relative
to the critical density according to the same definition as in
\S\ref{sec:lcdm} (there is no halo ``concentration'' in this
context).

Unlike the NFW halo with its cuspy $r^{-1}$ density centre, the LOG halo has
a constant-density core.
At larger radii, the density decreases as $r^{-2}$,
similar to the NFW model near $r=r_s$. This model allows us to
maximize the stellar contribution to the central mass, and to test
a ``minimal DM halo'' scenario. Similar models have been
successfully used to explain the dynamics of other galaxies of all
types (e.g. \citealt{1980MNRAS.193..189F,1991MNRAS.249..523B};
K+00; \citealt{2007MNRAS.382..657T,2008MNRAS.383.1343W}; DL+08;
\citealt{2010A&A...516A...4P}).

\subsection{Multi-component models: dynamical methods}\label{sec:dynmeth}

Our Jeans modelling approach has been extensively developed in
N+09, to which we refer the reader for the full description of the
equations adopted. Basically, in addition to the usual
second-order Jeans equations for the velocity dispersion profile,
we solve the fourth-order Jeans equations to constrain the LOSVD
with kurtosis data and reduce the systematic uncertainties linked
to the unknown orbital distribution (e.g.
\citealt{2001MNRAS.322..702M,lok02,lokmam03}). Although the
higher-order Jeans equations are not closed in general, one can
adopt a simple choice for the distribution function which makes
the problem tractable\footnote{We restrict ourselves here to
functions which can be constructed from the energy-dependent
distribution function by multiplying it by a function of angular
momentum $f(E, L) = f_0 (E) L^{-2 \beta}$ with $\beta =\rm const$
. This is a widely-used ansatz (\citealt{DF1}; \citealt{DF2};
\citealt{DF3}; \citealt{DF4}), which has the advantage of being
easy to integrate even though it does not generalize to the case
of $\beta=\beta(r)$ for the fourth-order moment.}. This
simplification is arbitrary (e.g. $\beta$ is assumed to be
constant with radius) and does restrict the generality of our
results, but the model is still more general than an assumption of
isotropy. In N+09 we demonstrated the utility of this approach for
assessing the presence of radial orbits in NGC 4494.

For the sake of clarity, we report in the following the basic
steps of our analysis (for more details, see also N+09):
\begin{enumerate}
\item Set up a multi-dimensional grid of model parameter space to
explore, including $\beta$ and the mass profile parameters
($\Upsilon_*, \rho_s, r_s$) or ($\Upsilon_*, v_0, r_0$).\\
\item For each model grid-point, solve the second- and fourth-order Jeans equations.\\
\item Project the internal velocity moments to $\sigma_{\rm los}$ and $\kappa_{\rm los}$.\\
\item Compute the $\chi^2$ statistic, defined as
\begin{equation}
\chi^2=\sum_{i=1}^{N_{\rm data}} \left[\frac{p^{\rm obs}_i-p^{\rm mod}_i}{\delta p^{\rm obs}_i}\right]^2 ,
\label{chi2}
\end{equation}
where $p^{\rm obs}_i$ are the observed data points ($\sigma_{\rm
los}$ and $\kappa_{\rm los}$), $p^{\rm mod}_i$ the model values,
and $\delta p^{\rm obs}_i$ the uncertainties on the observed
values, all at the radial position $R_i$. We fit the PN data
outside 60$''$ (where the spatial incompleteness due to the galaxy
background is more severe, see also \citealt{2001A&A...377..784N})
and the stellar data outside $10''$ (see \S\ref{sec:dynamics}).
\\
\item Find the best fit parameters minimizing the $\chi^2$.
In practice, we find that the VD is affected by both the mass and anisotropy profiles,
while the kurtosis is almost entirely driven by the anisotropy.
\end{enumerate}

\hspace{-2.5cm}
\begin{table*} \hspace{-2.5cm} \caption{Summary of
best-fit multi-component model parameters.} \label{tab:jeanssumm}
\scriptsize
\hspace{-1.5cm} \begin{tabular}{lcccccccccccc}\hline\hline
Model  & $\beta_5$$^1$ & \Ystar$^2$& log ${M_*}^3$ & $c_{\rm vir}$$^4$ & log ${M\vir}^5$& $f\vir$$^6$ & $f_{\rm DM,5}$$^7$ & $\Upsilon(\Re)$$^8$ & $\Upsilon_{B5}$$^9$ & $\Upsilon(R\vir)$$^{10}$ &$\nabla_{\ell} \Upsilon$$^{11}$ & $\chi^2/$ \\
 & & (\Ysol)  & (\Msun) &  & (\Msun) &  &  & (\Ysol)  & (\Ysol)  & (\Ysol)  & & d.o.f.$^{12}$ \\
\hline\hline
\multicolumn{12}{c}{No-DM model} \\
\hline
star iso & 0 & $7.5$ & 11.76 & -- & 11.76 & 0 & 0 & 7.5 & 7.5 & 7.5 & 0 & 123/36\\
\hline
\multicolumn{12}{c}{NFW model} \\
\hline
NFW iso & 0 & $6.4$ & 11.69$$ & 9$^{+8}_{-5}$ & 13.4$^{+0.4}_{-0.5}$& 54$^{+81}_{-36}$ & 0.7$^{+0.7}_{-0.4}$ &8$^{+2}_{-1}$ & 22$^{+14}_{-8}$ & 350 &0.47 & 28/45\\
NFW iso2 & 0 & $5.5$ & 11.62 & 12$^{+11}_{-6}$ & 13.3$^{+0.3}_{-0.5}$& 51$^{+82}_{-34}$ & 0.8$^{+0.9}_{-0.4}$  &7$^{+2}_{-1}$ & 23$^{+18}_{-10}$ & 286 &0.65 & 78/40\\
NFW+$\beta_0$& 0.2$\pm$0.1 & $5.5$ & 11.62 & 13$^{+10}_{-6}$ & 13.3$^{+0.3}_{-0.4}$ & 53$^{+59}_{-32}$ &  0.8$^{+0.7}_{-0.4}$  & 8$^{+2}_{-1}$ & 25$^{+14}_{-10}$ & 294 &0.72 & 23/44 \\
NFW+$\beta(r)$& 0.01$\pm$0.1 & $5.7$ & 11.64 & 14$^{+17}_{-8}$ & 13.1$^{+0.5}_{-0.6}$ & 32$^{+73}_{-25}$ &  0.7$^{+0.5}_{-0.4}$  & 8$^{+3}_{-2}$ & 22$^{+22}_{-11}$ & 183 &0.59 & 12/33 \\
NFW+AC+iso& 0 & $5.7$ & 11.64 & 8$^{+8}_{-5}$ & 13.3$^{+0.4}_{-0.6}$ & 45$^{+40}_{-25}$ &  0.7$^{+0.1}_{-0.3}$  & 7$^{+1}_{-1}$ & 17$^{+10}_{-10}$ & 261 &0.39 & 31/44 \\
NFW+AC+$\beta_0$& 0.30$\pm$0.15 & $5.5$ & 11.62 & 22$^{+17}_{-10}$ & 13.2$^{+0.5}_{-0.4}$ & 39$^{+43}_{-24}$ &  0.8$^{+1.0}_{-0.5}$  & 10$^{+2}_{-2}$ & 32$^{+18}_{-14}$ & 217 &1.0 & 40/44 \\
NFW+AC+$\beta(r)$&0.01$\pm$0.1 & $5.5$ & 11.62 & 7.5$^{+4.0}_{-3.0}$ & 13.4$^{+0.3}_{-0.4}$ & 66$^{+50}_{-37}$ &  0.7$^{+0.4}_{-0.3}$  & 7$^{+1}_{-1}$ & 18$^{+8}_{-6}$ & 368&0.44 & 15/33 \\
\hline \hline
\multicolumn{12}{c}{LOG model}\\
\hline
Model &  $\beta_5$$^1$ & \Ystar$^2$ & log ${M_*}$$^3$ & $v_0^{13}$  & log $M\vir^{5}$ & $r_0^{14}$ & $f_{\rm DM,5}$$^7$ & $\Upsilon(\Re)$$^8$ & $\Upsilon_{B5}$$^9$ & $\Upsilon(R\vir)$$^{10}$ &$\nabla_{\ell} \Upsilon$$^{11}$ & $\chi^2/$ \\
 & & (\Ysol) & (\Msun) & (kms$^{-1}$)  & ($M_\odot$) & (arcsec) &  & (\Ysol) & (\Ysol) & (\Ysol) & &  d.o.f.$^{12}$\\
\hline
LOG iso & 0 & $6.6$ & 11.70 & 456 & 13.66$^{+0.07}_{-0.08}$& 251 & 0.73$^{+0.05}_{-0.06}$ & 7.5$^{+0.5}_{-0.3}$ & 25$^{+6}_{-5}$ & 600 &0.57 & 25/45\\
LOG+$\beta_0$& 0.3$^{+0.1}_{-0.3}$ & 5.5 & 11.63 & 425 & 13.55$^{+0.12}_{-0.08}$& $190$ & 0.77$^{+0.06}_{-0.06}$ & 6.7$^{+1.3}_{-0.6}$ & 24$^{+8}_{-5}$ & 485 &0.73 & 27/44 \\
LOG+$\beta(r)$ &0.01$\pm$0.1& $6.0$ & 11.66& 412 & 13.52$^{+0.08}_{-0.09}$& $173$ & 0.75$^{+0.04}_{-0.05}$ & 7.4$^{+0.8}_{-0.5}$ & 24$^{+5}_{-4}$ & 440& 0.65 & 19/33\\
LOG+AC+iso & 0 & $6.3$ & 11.67& 443 & 13.62$^{+0.14}_{-0.13}$& 362 & 0.67$^{+0.11}_{-0.17}$ & 6.6$^{+0.7}_{-0.3}$ & 19$^{+11}_{-6}$ & 540 & 0.39 & 21/44\\
LOG+AC+$\beta_0$ & 0.3$^{+0.1}_{-0.3}$ & $5.5$ & 11.62& 419 & 13.54$^{+0.10}_{-0.07}$& $182$ & 0.77$^{+0.05}_{-0.05}$ & 6.9$^{+1.2}_{-0.6}$ & 24$^{+7}_{-4}$ & 465 & 0.72 & 28/44\\
LOG+AC+$\beta(r)$ &0.01$\pm$0.1& $5.5$ & 11.62& 403 & 13.48$^{+0.12}_{-0.09}$& $290$ & 0.70$^{+0.07}_{-0.08}$ & 6.0$^{+0.4}_{-0.2}$ & 18$^{+6}_{-4}$ & 414& 0.47 & 21/33\\
\hline \hline

\end{tabular}
\noindent{\smallskip}\\
\begin{minipage}{16.5cm}

NOTES: $1$) Anisotropy at the benchmark radius of 5\Re; $2$)
dynamical stellar mass-to-light ratio $M/L$, in $B$-band Solar
units: typical uncertainty is $\pm0.2$\Ysol; $3$) log of stellar
mass in solar units (uncertainties are of the order of 0.1 dex);
$4$) concentration parameter (see \S\ref{sec:lcdm}); $5$) log of
virial dark mass; $6$) ratio of total dark and luminous matter
within the virial radius, $f\vir=M_{\rm d}/M_*$ at $r_{\vir}$;
$7$) dark matter fraction, $f_{\rm DM}=M_{\rm d}/(M_{\rm d}+M_*)$
at  5\Re; $8$) dynamical $M/L$ at \Re; $9$) dynamical $M/L$ at
5\Re;
$10$) dynamical $M/L$ at the virial radius (uncertainties are of
the order of $50-70$\%); $11$) $M/L$ logarithmic gradient; $12$)
$\chi^2$ statistic (see text for details of data included);
$13$) asymptotic circular velocity (see Fig. \ref{fig:masscomp} for
uncertainties); $14$) halo core radius (see Fig. \ref{fig:masscomp}
for uncertainties).
\end{minipage}
\noindent{\smallskip}\\
\end{table*}

One interesting side-note is that given the assumptions of our
Jeans formalism, we showed in N+09 (Eqs. B10--B12) that if a
system has a constant dispersion profile, we can estimate its
internal anisotropy $\beta$ directly from the data without any
need for dynamical modelling. This is because the line-of-sight
kurtosis $\kappa$ is then a simple matter of projection effects
for a given $\beta$ and luminosity profile. Therefore at a radius
of $\sim 170''$, we estimate that NGC~4374 has an anisotropy of
$\beta \simeq -0.1^{+0.3}_{-0.4}$, i.e. it is near-isotropic.\\

The list of mass models we will explore in the following Sections
includes: 1) a no-DM case or self--consistent model where the
potential is given by the stellar mass only; 2) a NFW dark halo to
be tested against the $\Lambda$CDM predictions; 3) a core
logarithmic potential. The novelty of this analysis with respect
to N+09 and all other dynamical studies on individual ETGs is the
inclusion of the effect of the adiabatic contraction of the dark
halo, for both the DM halo models as above.

\subsection{Multi-component model results: no-DM case}\label{sec:dynres}

In \S\ref{sec:massinv} we have seen that for \N4374, a model with
a constant $\ML$ with radius is ruled out by the PN velocity dispersion data.
However, the pure-stellar potential ($\rho_s=0$
or $v_0=0$) is the minimal model that can be tried to fit the
dispersion and kurtosis data, allowing us to find the maximum
stellar content of the galaxy compatible with the inner data points.

The best-fit parameters of the model with an isotropic velocity
ellipsoid ($\beta=0$) are listed in Table~\ref{tab:jeanssumm}
together with the $\chi^2$ of the fit.

\begin{figure*}
\hspace{-0.7cm}
\begin{minipage}[b]{0.35\linewidth}
\includegraphics[scale=0.57]{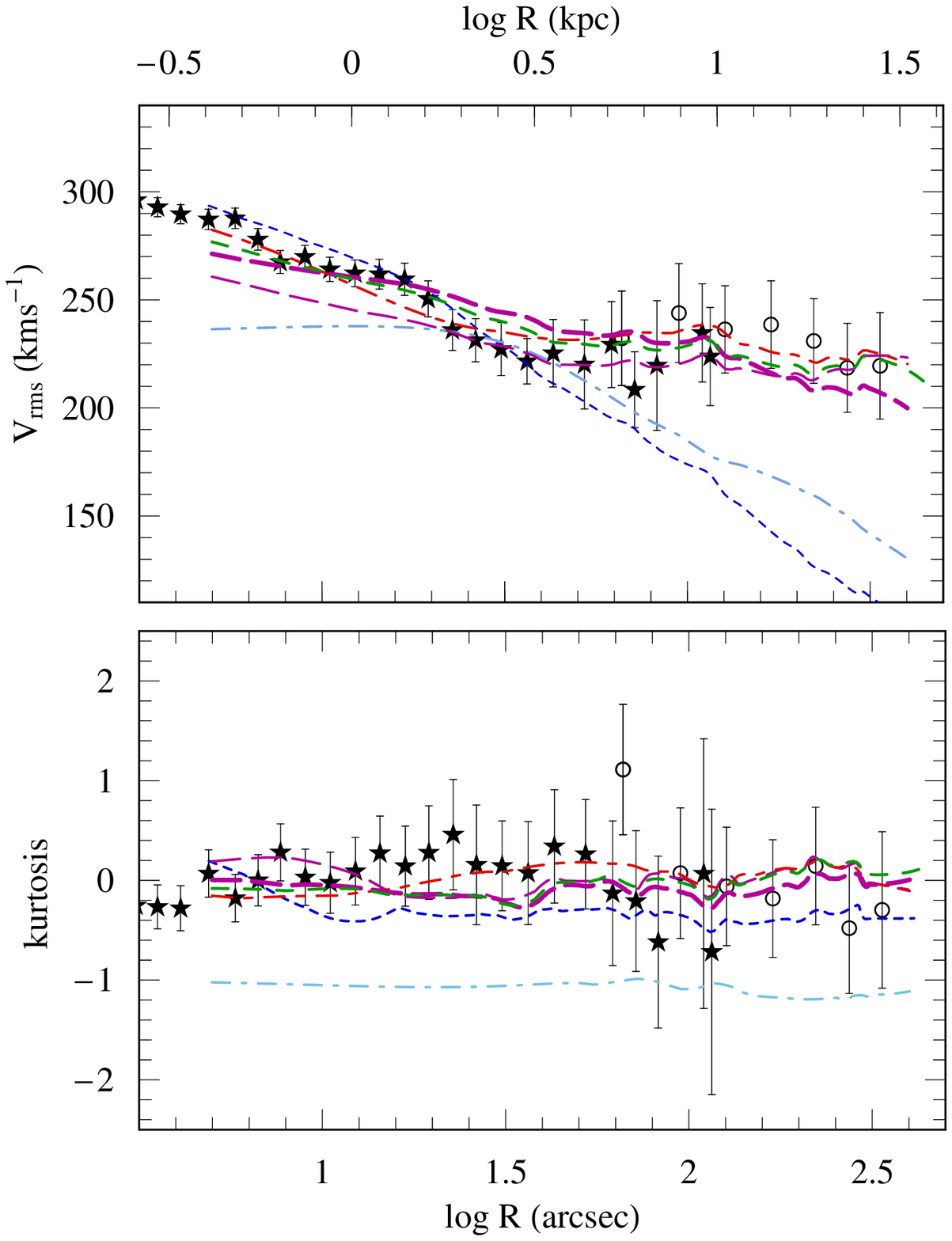}
\end{minipage}
\hspace{3.cm}
\begin{minipage}[t]{0.5\linewidth}
\vspace{-10.28cm}
\includegraphics[scale=0.47]{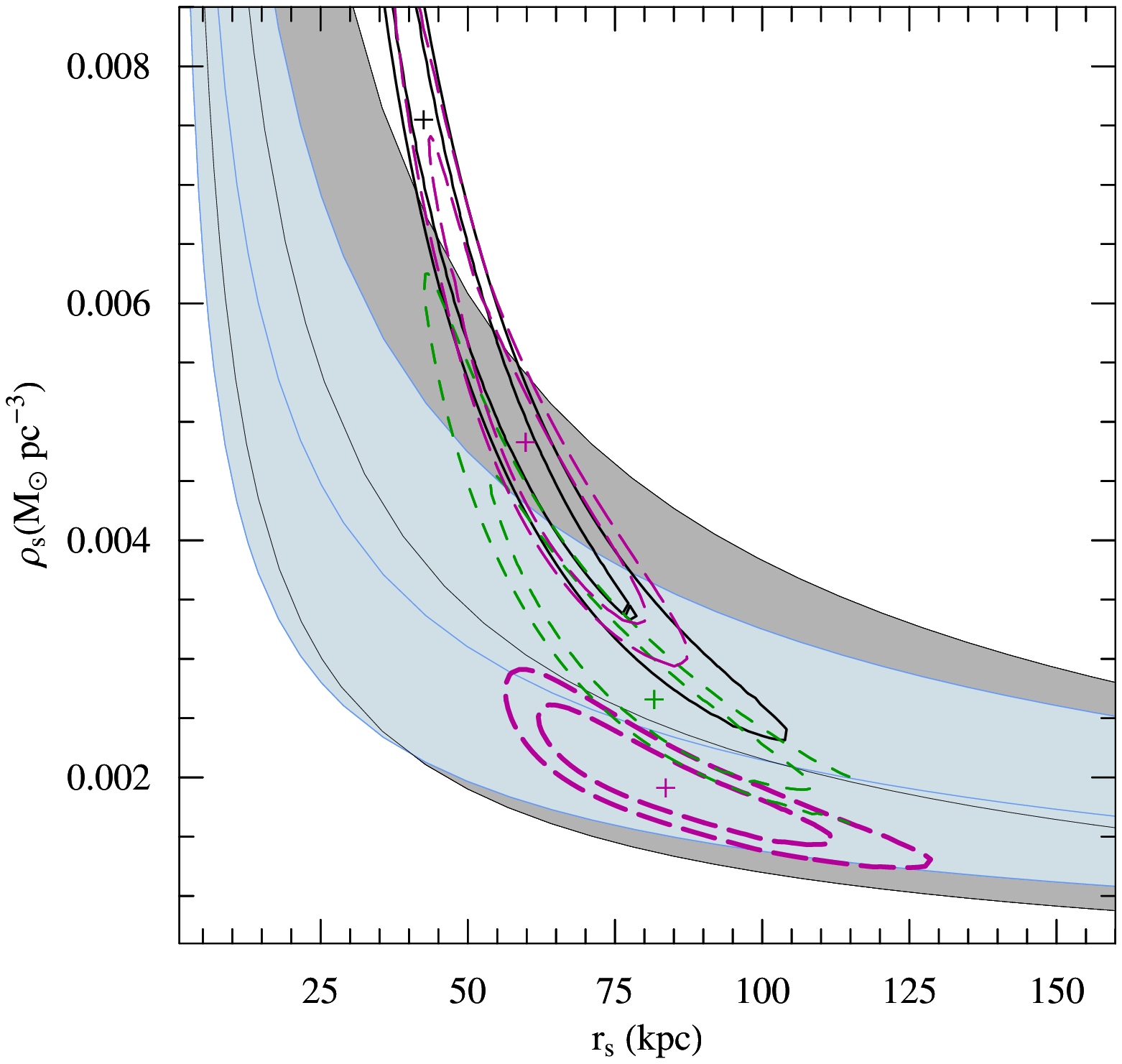}
\includegraphics[scale=0.6]{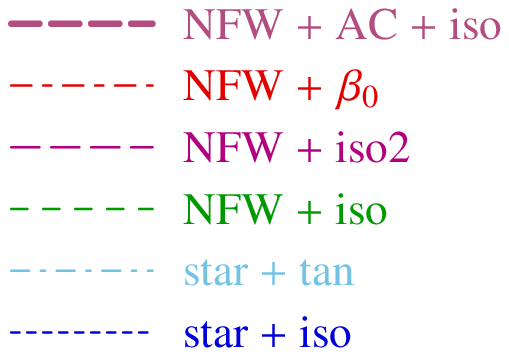}
\end{minipage}

\caption{Multi-component Jeans model fits to the NGC~4374
kinematics data. The stellar data are shown by star symbols, and
the PN data are open circles. The left panels show the projected
RMS velocity profiles (top) and the projected kurtosis (bottom),
the right panel the corresponding $1$ and $2~\sigma$ confidence
level of the $\rho_s-r_s$ parameters marginalized with respect to
$\Upsilon_*$ and $r_a$ (for the ``NFW+$\beta(r)$'' model). The
curves correspond to models as in the panel legend (except
``star+tan'' which is not a best-fit model). The shaded regions on
the right show the WMAP1 (gray) and WMAP5 (blue) expected region
for halo parameters. The ``NFW~$+ \beta(r)$'' model from
Fig.~\ref{fig:fig7} is plotted here for comparison with the
isotropic case and repeated in Fig.~\ref{fig:fig8}. See text for
details.} \label{fig:fig7}
\end{figure*}

Given the freedom to adjust $\Upsilon_*$, the model is able to fit
the VD in the central regions $(\lsim 2 \Re)$ with a best-fit
$\Upsilon_*=7.5$ ($V$ band). This value is consistent with the SSP
estimates based on the Salpeter IMF, and inconsistent with the
Kroupa IMF predictions at more than 1$\sigma$. We will come back
to this issue in the next Section, and note here that, despite the
higher $\Upsilon_*$, the no-DM model fails to reproduce the data
since the VD falls off too quickly in the outer regions
(Fig.~\ref{fig:fig7}, blue dotted line). The gap between the model
and the data cannot be removed even by assuming extremely negative
$\beta$ (see e.g. the cyan dot-dashed line for $\beta=-3\times
10^3$) or by adopting a shallower SB profile as allowed by the fit
errors in \S\ref{sec:spatdist}.


These Jeans models are not general enough to explore every
dynamical solution that is physically possible, but we judge that
the data/model differences are large enough to render a constant
$\ML$ model highly implausible.
We will next proceed with models allowing for the presence
of a DM halo to find out what halo parameters are most consistent
with the data for the two assumed DM profiles.

\subsection{Multi-component model results: NFW model}\label{sec:resNFW}
We next consider the NFW mass model (Section~\ref{sec:lcdm}) based
on $\Lambda$CDM expectations. We initially discuss the case with
orbital isotropy in \S\ref{sec:MLissue} and show that this matches
the data fairly well except near \Re\ (namely, $20''-100''$) where
the dispersion (kurtosis) is overestimated (underestimated) by the
Jeans models. In \S\ref{sec:MLissue} we explore a range of
constant and radially-varying $\beta$ profiles and conclude that a
significant radial anisotropy is ruled out at large galactocentric
distances, while the $\beta(r)$ profile as in Eq. \ref{eq:betar2}
provides the best match to the data at all radii. Finally we
include in our model the effect of adiabatic contraction in
\S\ref{sec:AC} and find that the higher central DM fraction
thereby generated allows the data to accommodate a smaller stellar
$M/L$, fully compatible with a Kroupa IMF.

\subsubsection{The isotropic model and the stellar M/L issue}\label{sec:MLissue}

We start by assuming isotropy, and find a best fit as shown in
Fig.~\ref{fig:fig7} (green dashed), with parameters again reported
in Table~\ref{tab:jeanssumm} (``NFW iso''). This solution is a fairly good match
to the data, for both the VD and kurtosis profile, which is a
further support for the absence of strong anisotropy in the
stellar orbital distribution. The best-fit $\Upsilon_* \sim 6.5
\Ysol$ is lower than the no-DM case because the central regions
contain significant amounts of DM
(see \S\ref{eq:betar2}), although it is still the stellar mass
that determines the main kinematical features inside $\sim 100''
\sim 1.2Re$. This stellar $\ML$ value is more consistent with a
Salpeter IMF than with Kroupa (to be addressed further in
%
\S\ref{sec:MLissue}).


The central NFW halo parameters of
$\rho_s=0.0030^{+0.0012}_{-0.0009}$~$ M_\odot$~pc$^{-3}$ and
$r_s=915''\pm200''=76\pm17$~kpc (see Fig. \ref{fig:fig7} which
shows the joint region of permitted values for $r_s$ and $\rho_s$,
marginalized over the other free parameters, $\Upsilon_*$)
correspond to a virial radius, mass, and concentration of $r_{\rm
vir}=770\pm70$~kpc, $M_{\rm vir}=(2.5^{+3.8}_{-1.7})\times 10^{13}
M_\odot$ and $c_{\rm vir}\sim9^{+8}_{-5}$.
These halo parameters are comfortably compatible with WMAP5
expectations (Eqs.~\ref{cMvir2} and \ref{eq:rho_rs_w5}), as well
as WMAP1 (modulo an IMF issue that we discuss below).
Looking carefully at the details of the DM halo solution, the VD
(kurtosis) data within $20''-100''$ ($1.3-2$ dex) are slightly
overestimated (underestimated) by the model, which might be either
an indication of 1) some degree of anisotropy or of 2) a mass
excess caused by a larger DM concentration not accounted for in
the NFW halo model.

\begin{figure*}
\hspace{-0.7cm}
\begin{minipage}[b]{0.35\linewidth}
\includegraphics[scale=0.57]{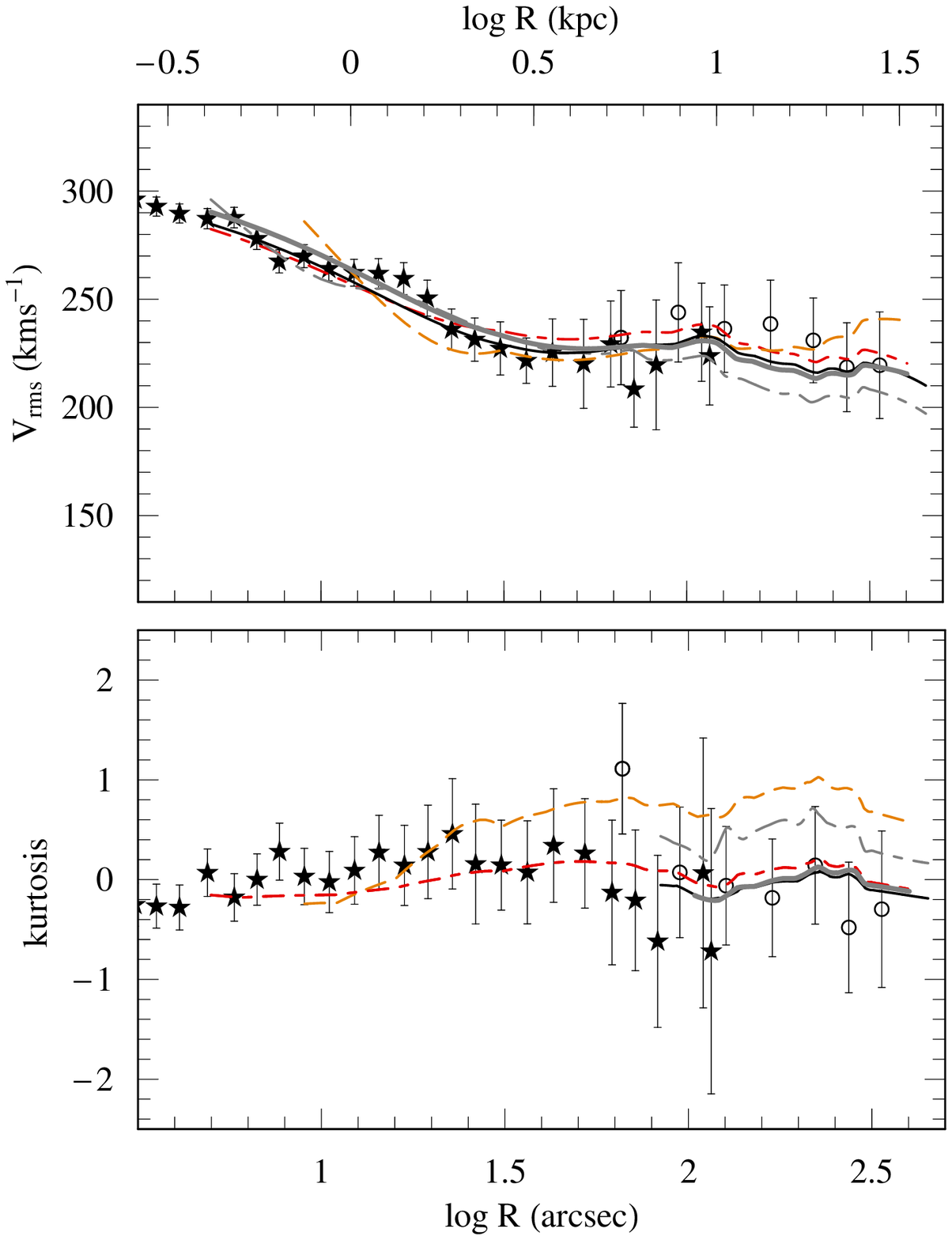}
\end{minipage}
\hspace{3cm}
\begin{minipage}[t]{0.5\linewidth}
\vspace{-10.28cm}
\includegraphics[scale=0.47]{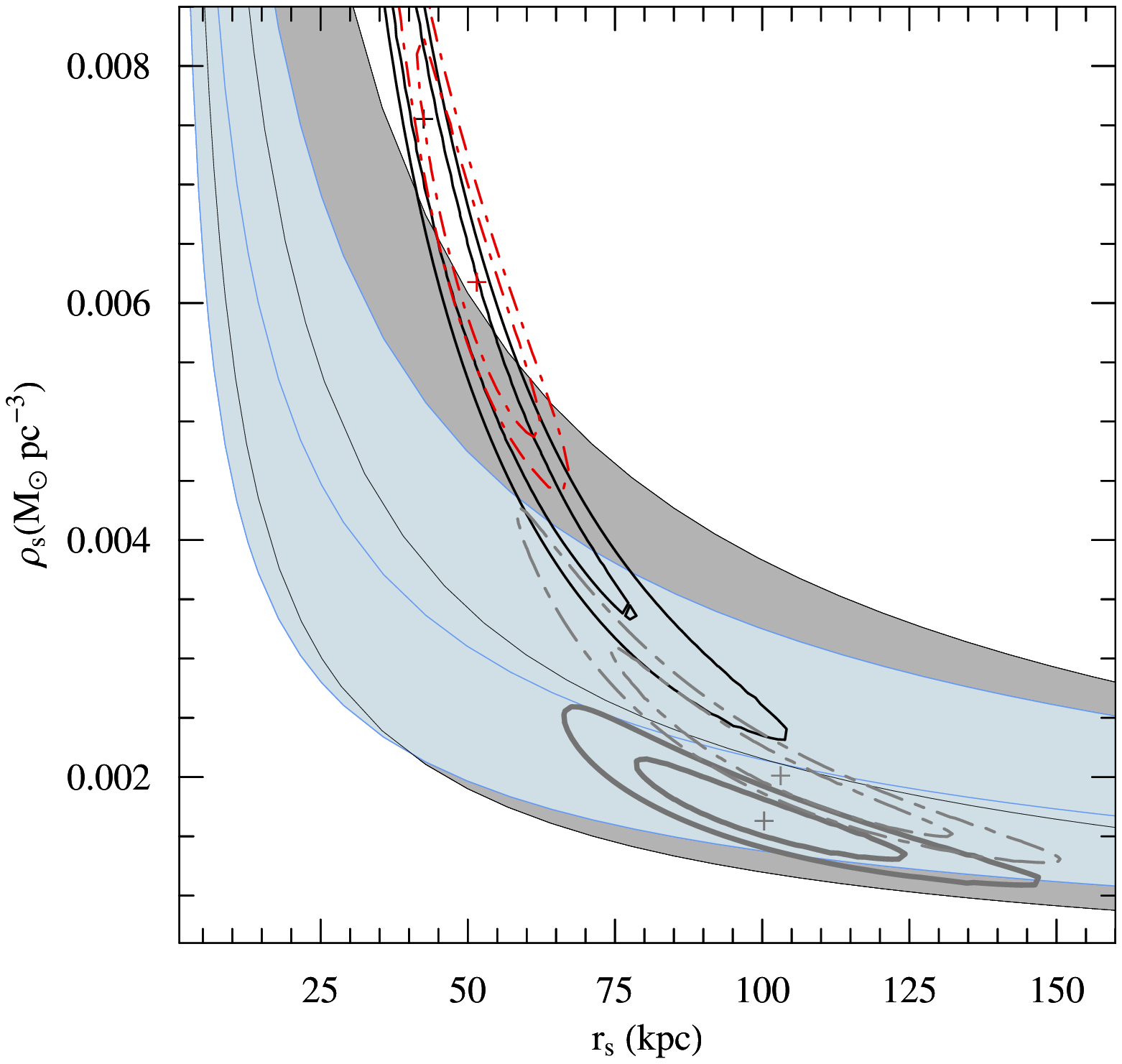}
\includegraphics[scale=0.6]{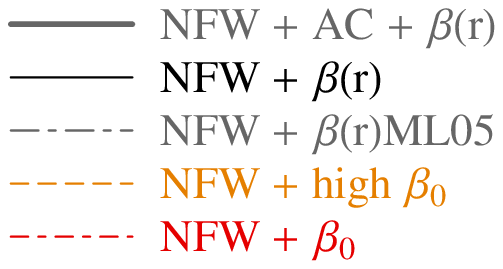}
\end{minipage}

\caption{As Fig.~\ref{fig:fig7}.
Confidence level of the $\rho_s-r_s$ parameters marginalized with
respect to $\Upsilon_*$ and $\beta_0$ or $r_a$ (except for
``NFW$+$ high $ \beta_0$'' which is not a best-fit model). The
``NFW$+ \beta(r)$'' model is repeated as overlap with
Fig.~\ref{fig:fig7} .} \label{fig:fig8}
\end{figure*}

Before we explore these two options, we will investigate further
the IMF issue mentioned above.

In the NFW dark halo model solutions discussed so far, the
best-fit $\Upsilon_*$ ($\sim 6.4$) is more comfortably consistent
with the stellar $\ML$ predicted by the population analysis
assuming a Salpeter IMF than a Kroupa IMF (see
\S\ref{sec:massinv}). Although this is not a strong argument for
preferring either IMF, we have tried to quantify the effect of
$\Upsilon_*$ on our result.

The high $\Upsilon_*$ is mainly driven by the fit to the central
data-points and the tendency of the $\chi^2$ procedure to favour
more minimal halo solutions. Since our simple Jeans models are not
designed to reproduce detailed kinematical structure as might be
present in the central regions, we lower the weight of the very
central VD and kurtosis data-points (i.e. data up to 30$''$,
$\sim$1.5 dex) in the $\chi^2$ minimization. In this case, more
centrally concentrated halo solutions can be made compatible with
the data\footnote{E.g., in Fig. \ref{fig:modelfitstwo} of
\S\ref{sec:massinv} a lower central $\ML$ is found (though for the
$\beta=+0.5$ case).}. Indeed, in Fig. \ref{fig:fig7} (thin purple
dashed line), we report the best fit obtained for the isotropic
assumption, where a lower stellar $\ML$ is needed,
$\Upsilon_*=5.5$, which implies a dark matter halo with
$\rho_s=0.0049^{+0.0021}_{-0.0013}$~$ M_\odot$~pc$^{-3}$ and
$r_s=720''\pm200''=60\pm17$~kpc corresponding to a virial radius,
mass and concentration of $r_{\rm vir}=720\pm30$~kpc, $M_{\rm
vir}\sim2.1\times 10^{13} M_\odot$ and $c_{\rm vir}\sim12$ (see
also ``NFW iso2'' solution in Table \ref{tab:jeanssumm}). In this
case, though, the halo concentration is higher than predicted for
WMAP5 parameters.

In Fig. \ref{fig:fig7} it is evident that this solution has a
shallow velocity dispersion profile at $R<25''\sim1.4$ dex which
is a poor match to the data and causes the high $\chi^2$ value for
the fit.
However, the gap can be filled either with the presence of some
(anticipated) degree of anisotropy in the central regions or by a
DM enhancement by an adiabatically contracted halo. In the
following, we will explore these two possibilities in turn.

\subsubsection{Models with orbital anisotropy}\label{sec:anis_prof}

A way to produce a
modelled steeper $\sigma_{\rm los}$ profile,
for a given slope of the intrinsic light
density profile, $j_*$, and velocity dispersion $\sigma_r^2$ (see
e.g. Eq. \ref{eq:masseq}), is with some degree of radial anisotropy
(see, e.g., Dekel et al. 2005).

We have started with a constant anisotropy from the very central
regions and the best-fit solution is found to accommodate a gentle
radial anisotropy ($\beta_0\sim0.2$) with a lower stellar $M/L$
($=$~5.5~$\Ysol$) that now agrees with a Kroupa IMF. The VD and
the kurtosis are at last reproduced well at all fitted radii
(Fig.~\ref{fig:fig8}, red dot-dashed line), which is reflected in
an improved $\chi^2$ value in Table~\ref{tab:jeanssumm} (``NFW$+\beta_0$'').

The halo concentration for this solution is fairly high, and
just consistent with the WMAP5 expectations at the $\sim$~1$\sigma$ level.

We remark here that the constant anisotropy solution provides a
compromise model dispersion curve among regions which might have
different orbital structures. For this reason we decided to test
also the case of a radially varying $\beta(r)$ even though our dynamical procedure
is not explicitly designed for this. As done in N+09, we will use
the kurtosis data to constrain $\beta$ in the outer
regions where the anisotropy may be approximately constant.

Following the approach of \S\ref{sec:massinv} we use the
$\beta(r)$ as in Eq. \ref{eq:betar2}. The best-fit model is shown
in Fig. \ref{fig:fig8} (black line) and the parameters are
reported in Table \ref{tab:jeanssumm} [``NFW$+\beta(r)$'']. The anisotropy radius $r_a$
turned out to be very close to the one estimated with the
pseudo-inversion procedure ($r_a=33''$). The match in the central
regions is remarkably good also for the low $\Upsilon_*$, while in
the outer regions the model tracks the isotropic case (see left
panel of Fig. \ref{fig:fig7} for a direct comparison), and the
halo concentration is again somewhat on the high side
(see Fig. \ref{fig:fig8}, right panel).

We have also checked that outside 100$''$ radial anisotropy is
disfavoured: even when forcing the $\Upsilon_*$ to lower values (we
tried different values down to $\Upsilon_*=5$), in order to
allow for more radial anisotropy, the match to the outer data,
especially the kurtosis, was poor (see dashed orange line).
This result is somewhat surprising since predictions from galaxy
formation simulations generally show a significant degree of
radial anisotropy (see e.g. M{\L}05 and references therein), which
has been confirmed by dynamical analysis in the case of a few galaxies
(R+03; N+09; DL+09; but see \citealt{2010ApJ...716..370F}).
Indeed, we have used directly the M{\L}05 expression (see Eq.
\ref{eq:MLbeta}) in modelling our data and found that the fit to
both the VD and the kurtosis was possible only with a too small
$r_a(\sim6'')$, which is completely inconsistent with the values
found by \citet[ i.e. 1.4 \Re; see Fig.~\ref{fig:fig8} gray
dot-dashed line]{2005MNRAS.363..705M}. Fixing $r_a$ to the
expected value, the fit was possible only with a larger
$\Upsilon_*\sim 6.5$. In either case, though, a much poorer
significance of the fit than the one given by our preferred $\beta
(r)$ profile (Eq. \ref{eq:betar2}) was found.

In summary, our exploration of the NFW models indicates that halo
parameters corresponding to WMAP5 expectations are compatible with
the data.  The agreement is better for a Salpeter IMF, with the
concentration becoming somewhat high for a Kroupa IMF.
The near-isotropic orbital distribution that we infer is
at odds with standard predictions for radial orbits.
However, as discussed in the Appendix, there are some uncertainties
in the classification of velocity outliers, such that we cannot yet
claim the isotropy conclusion as robust.

\subsubsection{Effect of adiabatic contraction}\label{sec:AC}
The baryonic collapse occurring during galaxy assembly is one of
the physical process that can shape the central DM distribution in
a way different from the predictions of the dark matter only
N-body simulations. Given a dark matter halo distribution with the
properties predicted by such simulations, the (collisional)
collapsing gas can exert a dynamical drag on the DM particles and
produce a more concentrated final DM density profile (see e.g.
Blumenthal et al. 1986). The net effect is a larger central DM fraction
and consequently a lower stellar mass contribution (i.e. a lower
$\Upsilon_*$) to the total mass in the central regions
(for fixed dynamical $\ML$ and halo parameters).

\begin{figure*}
\hspace{-0.7cm}
\begin{minipage}[b]{0.35\linewidth}
\includegraphics[scale=0.57]{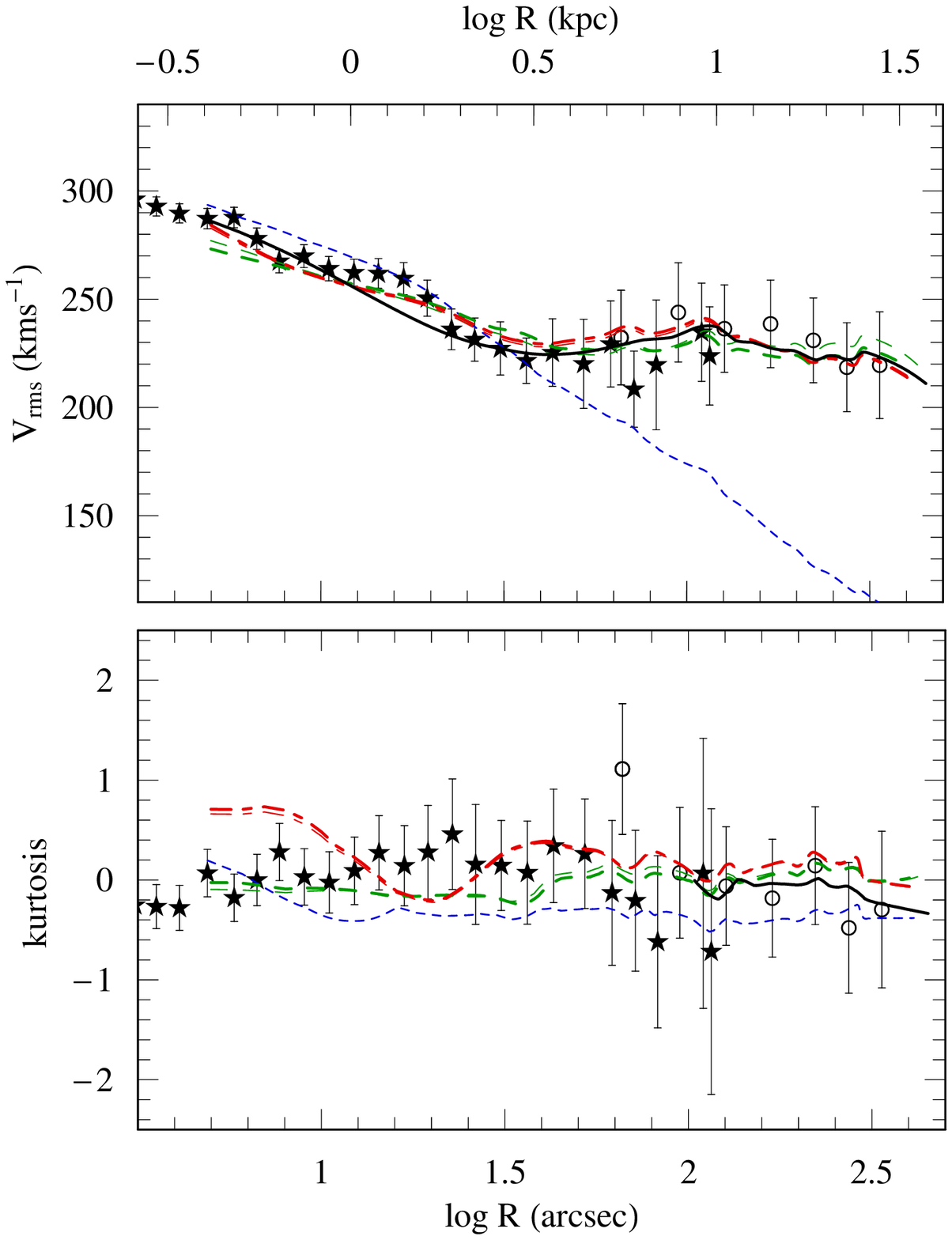}
\end{minipage}
\hspace{3cm}
\begin{minipage}[t]{0.5\linewidth}
\vspace{-10.28cm}
\includegraphics[scale=0.47]{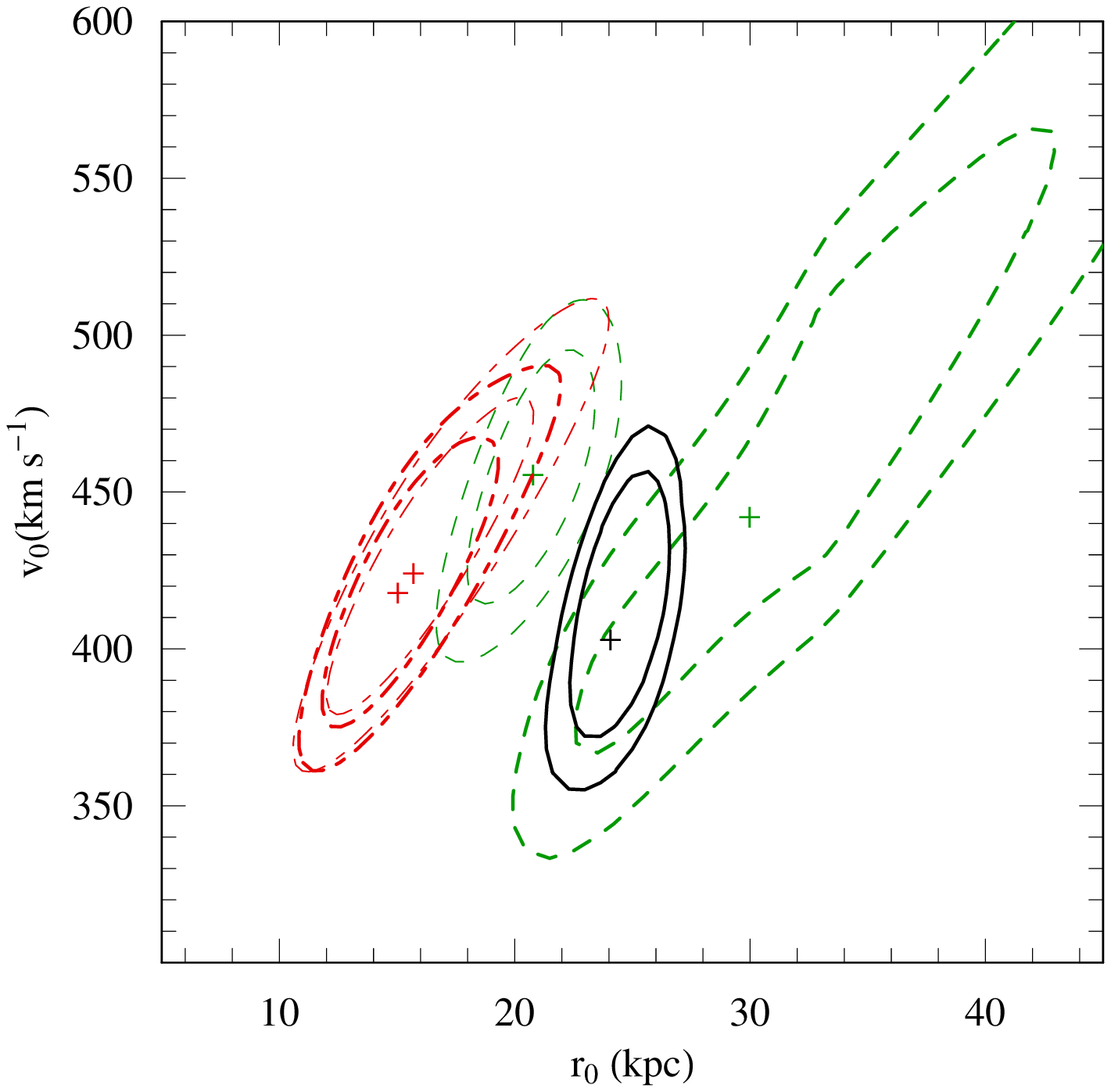}
\includegraphics[scale=0.6]{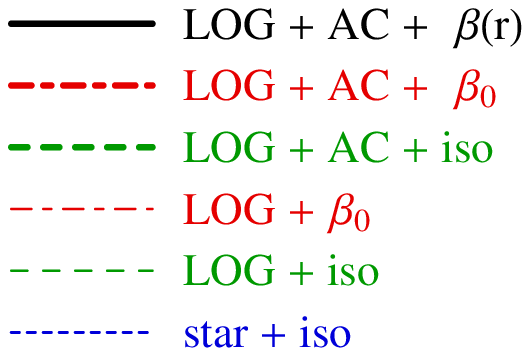}
\end{minipage}

\caption{As Figs.~\ref{fig:fig7} and \ref{fig:fig8}, with LOG
models. The right-hand panel shows the corresponding $1$ and
$2~\sigma$ confidence level of the $v_0-r_0$ parameters
marginalized with respect to $\Upsilon_*$ and $\beta$ parameters
(when available). The curves correspond to models as in the panel
legends. See text for details.} \label{fig:fig10}
\end{figure*}

This process can be described analytically by an adiabatic
contraction (AC hereafter; \citealt{1986ApJ...301...27B};
\citealt[G+04 hereafter]{2004ApJ...616...16G})
of the dark halo. Since there is not yet a final consensus on
the effectiveness and accuracy of the descriptions on the
market (see e.g.
\citealt{2010MNRAS.402..776P,2010MNRAS.405.2161D,2010MNRAS.406..922T}),
we decided to use the recipe from G+04.
The G+04 model produces a weaker effect on the final DM distribution
than the original Blumenthal recipe, and appears closer to the results obtained in the
cosmological simulations including the baryon physics.

A critical evaluation of the baryonic processes is beyond the
purpose of this analysis, where we only intend to check whether
including an analytical recipe for AC in our Jeans analysis would
provide a viable explanation to reconcile the estimated
$\Upsilon_*$ derived from our analysis and the stellar population
models. Furthermore, to our knowledge, the use of the AC in
detailed Jeans modelling of the velocity dispersion profile of an
elliptical galaxy has not been attempted before, so we consider
this an interesting exercise even though the AC recipe might be
not optimal.

For this purpose, in our equations
the total mass generating the potential $\Phi=G M(r)/r$ is given
by considering as an adiabatic invariant the quantity
\begin{equation}
M(\bar{r})r={\rm const} \label{eq:gnedin1}
\end{equation}
where $\bar{x}=Ax^w$ and $x=r/r_{\rm vir}$. By calibrating Eq.
\ref{eq:gnedin1} to collisional N-body simulations, G+04 have fixed
$A=0.85$ and $w=0.8$. The contracted DM mass distribution has been
derived by solving the equation
\begin{equation}
[M_{\rm tot}(\bar{r})]r=[M_{\rm DM}(\bar{r})+M_{\rm *}(r_f)]r_f
\label{eq:gnedin2}
\end{equation}
where $M_{\rm tot}=M_{\rm DM}+M_*$, and $M_{\rm DM}$ and $M_*$ are
the final dark and stellar mass respectively (initially assumed
to have the same spatial distribution). The model
results are shown in Fig. \ref{fig:fig8} and the model parameters
in Table \ref{tab:jeanssumm} [``NFW+AC+iso, $+\beta_0$, $+\beta(r)$''].
There are two main remarks that we can derive from these results.

First, since the effect of the AC is to drag more DM into the
central regions, the $\Upsilon_*$ turns out to be smaller than in
the no-AC case. For the isotropic case we obtain
$\Upsilon_*=$~5.7~$\Ysol$ (see Fig. \ref{fig:fig7}, tick purple
dashed line), but if we again include $\beta(r)$ as in Eq.
\ref{eq:betar2}, the best fit is found for
$\Upsilon_*=$~5.5~$\Ysol$ and $r_a=33''$. The goodness of these
fits is slightly worse than, but similar to, the uncontracted NFW
models
(see Table \ref{tab:jeanssumm}) with the model curves looking very
similar to the eye (see Fig.~\ref{fig:fig8}, tick gray
line)\footnote{The model with constant anisotropy and AC yielded a
relatively poor fit, and a very high halo concentration
(Table~\ref{tab:jeanssumm}).}.

Second, the (pre-contraction) dark halo parameters turn out to
be in very good agreement with $\Lambda$CDM.
E.g., for the anisotropic model, the NFW dark halo turns out to have $c_{\rm
vir}=7.5$ which matches the WMAP5 expectation ($c_{\rm
WMAP5}\sim7$ for $\log \Mvir=13.4$).

When forcing the fit to a lower $\Upsilon_*=$~5~$\Ysol$, the halo
parameters change slightly: the best fit is $c_{\rm vir}\sim9.1$ and
$\log \Mvir=13.5$, which is higher than the typical prediction but still
consistent with the scatter.

This is one of the most notable results of this paper:
for the first time using stellar kinematics extended out to
$\sim$~5~\Re, it has been demonstrated that the dark matter
content of a giant elliptical galaxy may be compatible with
$\Lambda$CDM.

\begin{figure*}
\hspace{-1.2cm}
\epsfig{file=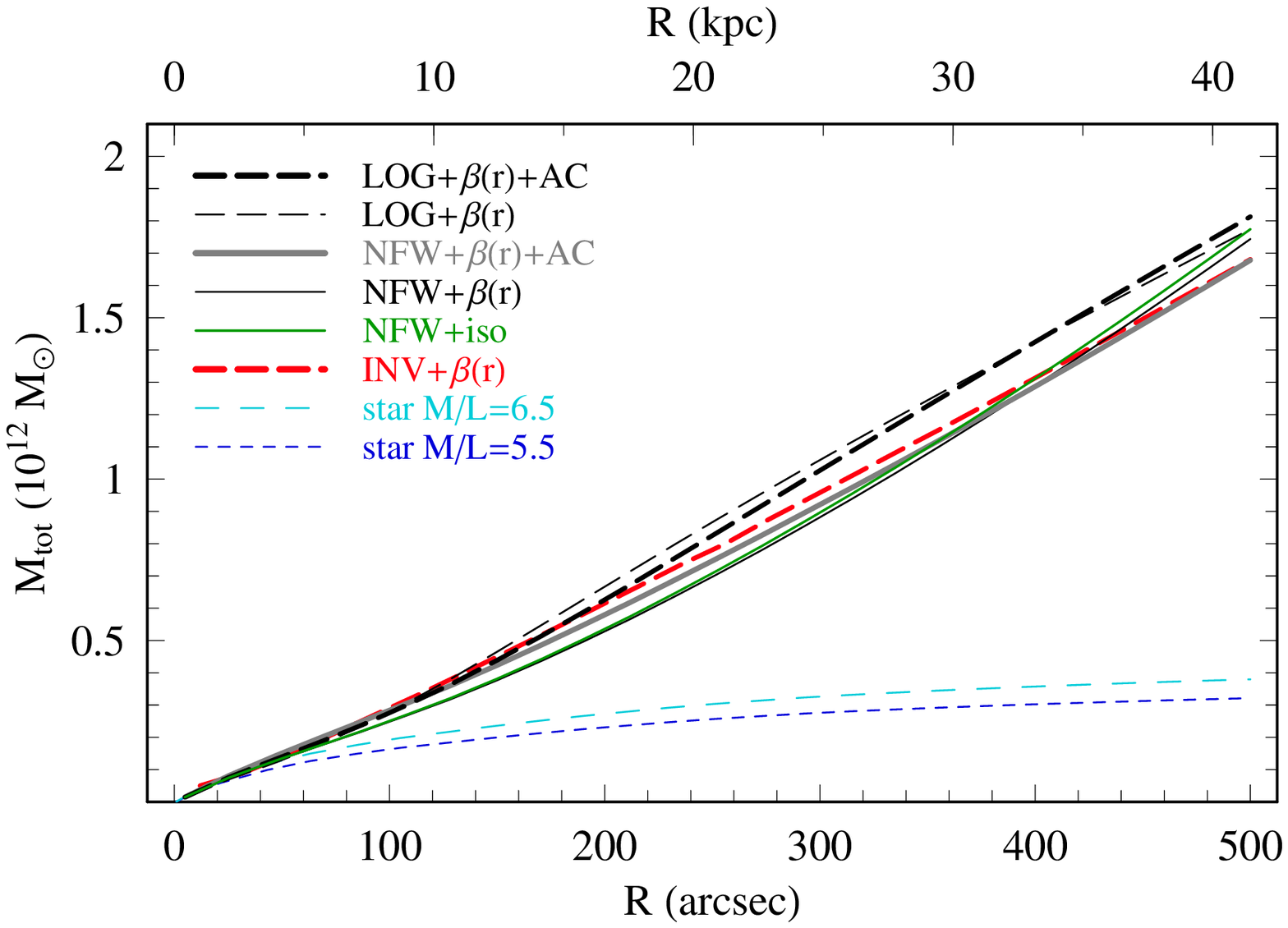,width=9.9cm}
\hspace{-1.3cm}
\epsfig{file=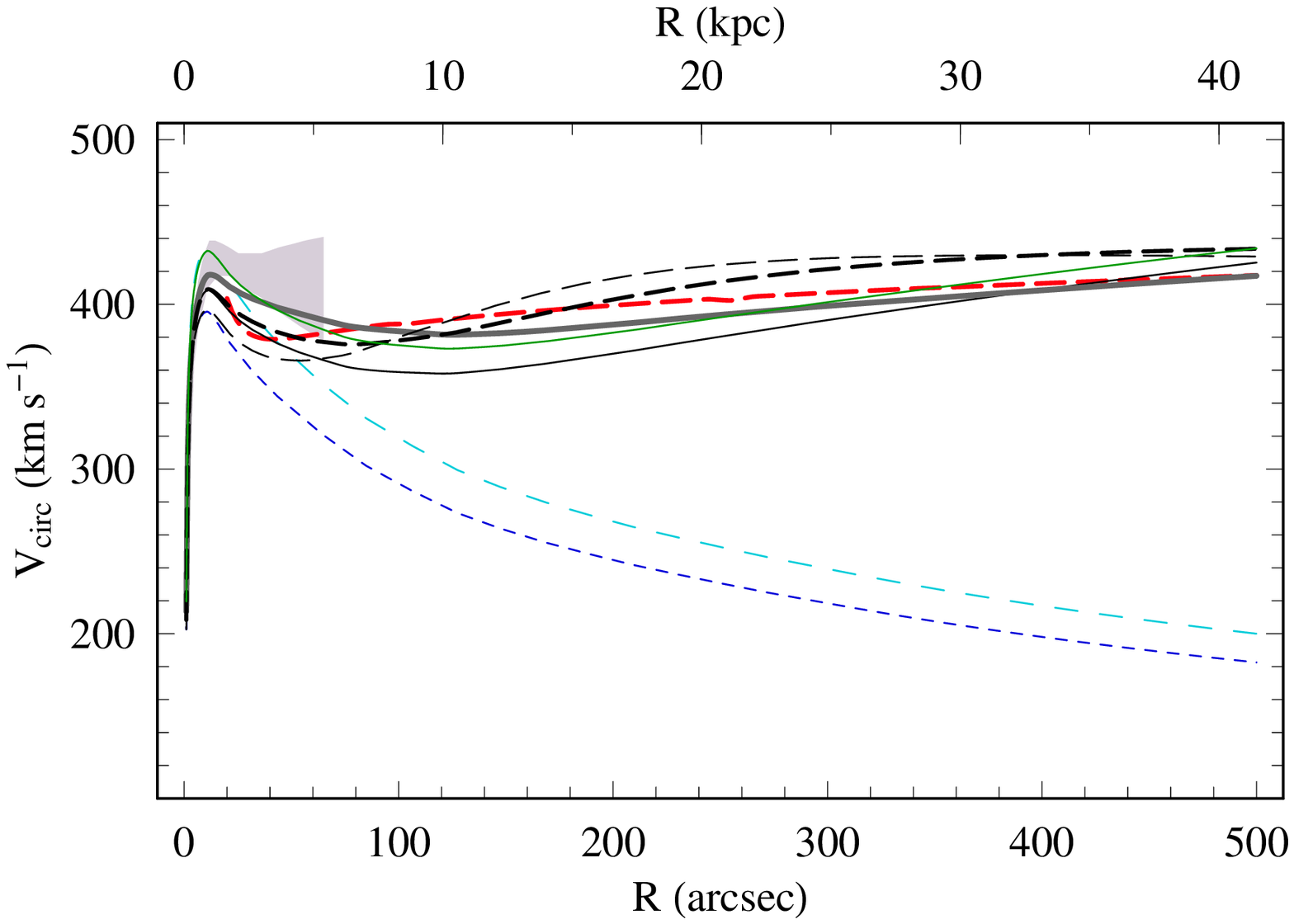,width=9.9cm}
\caption{Radial mass distribution of NGC~4374. The left panel
shows the cumulative mass, and the right panel shows the circular
velocity profile. Model curves from this work are as in the
legend. We also show the \vc\ profile from
\citet{2000A&AS..144...53K} (shaded area includes the variance of
their models).} \label{fig:masscomp}
\end{figure*}

\subsection{Multi-component model results: LOG model}\label{sec:resLOG}

We next carry out the model sequence for the LOG mass model (Section~\ref{sec:LOG}),
with results shown in Fig.~\ref{fig:fig10} and
Table~\ref{tab:jeanssumm}.
\subsubsection{The isotropic model}
For the isotropic case,
the LOG model can fit the data better than the NFW model in the
central regions and equivalently well in the outer regions (see
Fig. \ref{fig:fig10}, thin green dashed line). This is because the
LOG potential has an internal core with little DM contribution in
the central regions. In this case we find also a large stellar
mass-to-light ratio, $\Upsilon_*=$~6.6~$\Ysol$, which is more
compatible with a Salpeter IMF than Kroupa.
A massive DM halo is required outside $\sim$~100$''$
(mean $v_0\sim450\kms$; see Table \ref{tab:jeanssumm}, ``LOG iso''),
consistently with the pseudo-mass inversion analysis and the NFW
solution (see Fig. \ref{fig:masscomp}).

\subsubsection{Models with orbital anisotropy}
Adopting a constant non-zero anisotropy ($\beta_0=0.3$) allows for
a Kroupa-compatible
$\Upsilon_*=$~5.5~$\Ysol$ (the same as found using the NFW$+$AC model).
However, the fit is poorer (see Table
\ref{tab:jeanssumm}, ``LOG$+\beta_0$''), in particular at very small radii
(even though these are penalized in our model) and
owing to the higher estimates of the kurtosis at $R>100''$(=2
dex), as shown in Fig. \ref{fig:fig10} (thin red dot-dashed line).

We have checked if larger $\beta_0$ could be consistent with the
data at large radii and found that once $M/L_*$ and $r_c$ are
fixed, there is a degeneracy between the \vc\ and the $\beta_0$
values: a reasonable fit to the data is obtained for $\vc=410\kms$
and $\beta_0=0.1$ and $\vc=470\kms$ and $\beta_0=0.5$ with
$\ML_*=$~6~$\Ysol$ and $r_c=25$ kpc. Once again, the kurtosis
helps to put constraints on the allowed $\beta_0$: the
$\chi^2/$d.o.f. calculated over only the model versus observed
kurtosis profiles is much smaller for $\beta_0=0.1$ ($\sim9/20$)
than for $\beta_0=0.3$ ($\sim12/20$) and $\beta_0=0.5$
($\sim22/20$), which is a final demonstration that strong
anisotropy can be excluded at large radii.

Finally, we have adopted the $\beta(r)$ as in Eq. \ref{eq:betar2}.
The best-fit model is not showed (but almost identical to the one
with AC as in \S\ref{sec:LOGAC}), while parameters are reported in
Table \ref{tab:jeanssumm} [``LOG$+\beta(r)$'']. The anisotropy
radius $r_a$ is slightly larger that the one estimated with the
pseudo-inversion procedure and NFW ($r_a=45''$), although the
$\beta(r)$ profile turns out to be almost unaltered. The
$\Upsilon_*=6\Ysol$ is closer to the isotropic case, since this is
mainly constrained by the central regions which are almost
isotropic according to the Eq. \ref{eq:betar2}.

\subsubsection{Adiabatic Contraction}\label{sec:LOGAC}
For completeness, we have modelled the effects of a hypothetical
AC on the LOG halo. Because of the non-cuspy nature of the initial
halo, AC turns out to have only a weak affect, and does not change
any of the above conclusions. Model curves are almost
indistinguishable from the ones with no--AC as shown in
Fig.~\ref{fig:fig10} (green thick dashed line: isotropy; red thick
dot-dashed line: constant anisotropy) as a consequence of best-fit
parameters very close to the ones obtained for no--AC [Table
\ref{tab:jeanssumm} ``LOG+AC+iso, $+\beta_0$, $+\beta(r)$'', and
confidence contours in Fig.~\ref{fig:fig10}].

Finally,  the simultaneous use of the $\beta(r)$ anisotropy as in
Eq. \ref{eq:betar2} and the AC allowed the best fit to the data
(black thick line) as for the NFW case. For the LOG potential the
stellar \ML\ turned out to be $\Ystar=5.5 \Ysol$ and $v_0=403\kms$
(see Table~\ref{tab:jeanssumm}), and the anisotropy radius turned
out to be very similar to the NFW models ($r_a=35''$). Once again
the AC seemed to be a crucial ingredient to alleviate the problem
of the stellar \ML\ problem by naturally accommodating a
Kroupa-like \Ystar .

\subsection{Summarizing the best halo models: mass profiles and circular
velocities}\label{sec:mass_vcirc}

Before we discuss the implications of the best-fit solutions from
the previous sections, we summarize the models which we consider
more physically meaningful. As seen in Table \ref{tab:jeanssumm}
and discussed earlier, most of the models presented are
statistically good fits (e.g. the reduced
$\tilde{\chi}^2=\chi^2/$d.o.f. is almost everywhere $<1$), but
some of the models were incompatible with related theoretical
predictions.

For instance, the no-AC
models ``NFW+$\beta_0$'' and ``NFW+$\beta(r)$'' have
$\tilde{\chi}^2=0.5,0.4$ respectively, but the implied halo concentrations
are improbable given the $\Lambda$CDM expectations.
Also ``NFW+iso'' has a rather small
$\tilde{\chi}^2=0.6$ and a fairly $\Lambda$CDM-like halo, but the
large Salpeter-like $\Ysol$ makes this solution unfavourable.
On the other hand the model ``NFW+AC+$\beta(r)$'' has a
$\tilde{\chi}^2=0.45$ and is fully consistent with both
$\Lambda$CDM concentrations and a Kroupa IMF, and so is considered
our best reference model. For similar reasons, the favoured LOG
models are the ``LOG+AC+$\beta(r)$'', ``LOG+$\beta_0$'' and
``LOG+AC+$\beta_0$''---all having $\tilde{\chi}^2\sim0.65$ and a
$\Ysol$ compatible with a Kroupa IMF.

Going to the comparison among the different potentials compatible
with the stellar kinematics, in Fig. \ref{fig:masscomp} we plot
the mass profiles of some of these model solutions in order to
gain a general sense of
the different halo solutions accommodated by the data.

Considering the mass profiles, $M(r)$, for the different models
discussed above, the DM halo models (NFW and LOG) are
very different from the no-DM case, with the
$v_{\rm c}$ remaining much flatter with radius than the stellar
model.

The mass profile at 5~\Re{} ($\sim 30$ kpc) is remarkably similar
for the NFW and LOG models, demonstrating that this quantity is
well constrained by the data, independently of the details of the
mass models.

Despite the uncertainties, for the NFW case the mass profiles as
well as the $v_{\rm circ}$ profiles differ in the very central
regions when comparing the un-contracted solutions and the
contracted haloes.
The relative normalization between the stellar and halo masses changes due to
the higher dark mass allowed by the AC for a given halo
concentration before the contraction.
For the LOG model,
$\Ystar$ seems to be more degenerate with the $\beta$ value in the
central regions (in the sense that higher $\beta$ would allow
smaller $\Ystar$, see \S\ref{sec:LOG}). Overall, the
$v_{\rm circ}$ profiles (Fig. \ref{fig:masscomp}) turn out to be fairly
similar among the different models up to the last datapoint ($\sim
340''$), and beyond, if the profiles are extrapolated more deeply
into the halo regions. Furthermore, the mass profiles are remarkably
similar to the results of the pseudo-inversion
method (see Fig. \ref{fig:modelfitstwo}).

Finally, in Fig. \ref{fig:masscomp} we compare our results with
the $v_{\rm circ}$ profile from K+00,
which is based on long-slit data extending out to $\sim70''$.
Focusing on our LOG~$+\beta(r)$ solution which is the most
equivalent to theirs, our results are identical in the very
central regions, with a slight discrepancy at larger radii. Note
that the $v_{\rm circ}$ from K+00 extrapolated to 300$''$ (Fig.17
in \citealt{2001AJ....121.1936G})  is significantly lower than our
new profile based on more extended data and models.

The asymptotic run of all the model curves in Fig.
\ref{fig:masscomp} is remarkably tight which means that at
intermediate scales (of the order of the $r_s$ scale of the NFW haloes)
the overall galaxy mass is quite well constrained and the scatter
introduced by the halo models and the allowed anisotropy is small.
However, an important cross-check would be to verify how these
models might differ around the virial radius, where the NFW and LOG
profiles are expected to differ significantly (although the
extrapolated $M_{\rm vir}$ values in Table~\ref{tab:jeanssumm} do
not differ much).

\section{Discussion}\label{sec:discuss}

The dynamical solutions for the bright elliptical NGC 4374 all clearly indicate that
this galaxy is surrounded by a massive DM halo.
DM haloes were also found in four ordinary ETGs studied using PNe
(not all of these studies using PN constraints):
NGC~3379 (R+03; DL+09; \citealt{2009MNRAS.398..561W}); NGC~4697 (DL+08);
NGC~4494 (N+09; \citealt{2010arXiv1007.5200R});
NGC~821 \citep{2009MNRAS.398..561W,2010ApJ...716..370F}.
Apart from alternative gravity theories (e.g. \citealt{2007A&A...476L...1T}),
it seems clear that elliptical galaxies in general contain DM,
and the question is how the DM profiles compare in detail to predictions.

Some of the galaxies above were modelled with NFW haloes and some with LOG haloes,
while NGC~4374 is the first of these cases where {\it both} were tried.
Unfortunately, we were not able to discriminate between the two models, given
the limitations of the simple Jeans approach which cannot fit the observations in
great detail and requires somewhat arbitrary weighting of the data points.
Interestingly, the two models do seem to prefer different \Ystar\ values,
corresponding to Kroupa and Salpeter IMFs for the NFW and LOG haloes, respectively.
Adopting a prior on the IMF may then provide more information about the DM profile,
and vice-versa.
More detailed modelling may also be able to discriminate between these haloes
on the basis of dynamics alone:
e.g. with much less extensive data in a sample of ETGs, but using Schwarzschild modelling,
\cite{2007MNRAS.382..657T} found some suggestions that LOG haloes were preferred over NFW.

\begin{figure}
\centering \hspace{-1cm}
\epsfig{file=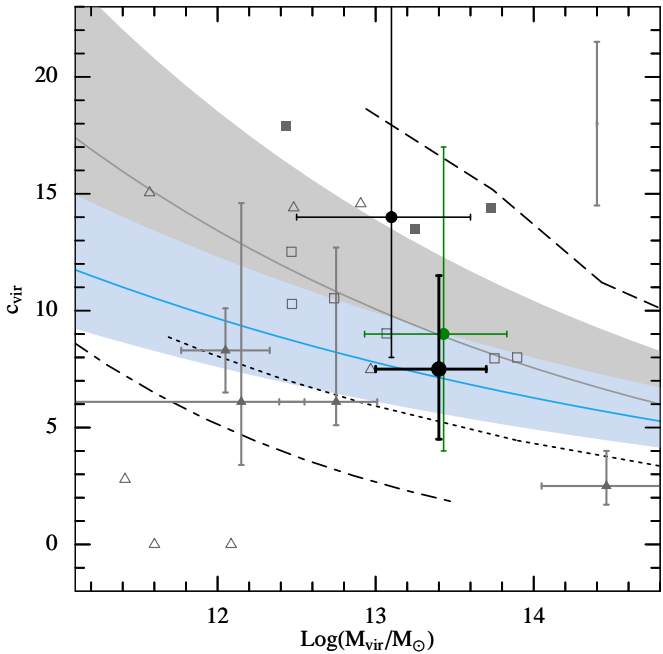,width=8.8cm} \caption{Dark
matter halo virial mass and concentration parameters. Several
reference solutions for NGC~4374 ({\it large filled circles}) are
plotted as well as other data taken from N+09. The blue and gray
curves with surrounding
shaded regions are the WMAP5 and WMAP1 predictions, respectively.
The green small dot with error bars is the ``NFW iso'' solution
(see Table \ref{tab:jeanssumm}; the stellar $M/L$ is consistent
with a Salpeter IMF), the black small dot is ``NFW$+\beta(r)$''
(corresponding to a Kroupa IMF),
and the big black dot to our
favoured model ``NFW+AC+$\beta(r)$''. From N+09: {\it Triangles}
and {\it boxes} mark fast-rotator and slow-rotator ETGs,
respectively. The {\it small filled symbols} mark detailed ETG
dynamical results using PNe and GCs (including error bars, where
available). The {\it open symbols} show the dynamics-based ETG
results from N+05, with error bars in the upper right corner
showing the typical uncertainties. The {\it dashed line} shows the
mean result for X-ray bright groups and clusters, the {\it
dot-dashed curve} is an inference for late-type galaxies, and the
{\it dotted curve} is the trend from weak lensing of all types of
galaxies and groups (see N+09 for details). }
\label{fig:confb1}
\end{figure}

Adopting the NFW halo model for the time being, it is important to test the
inferred halo parameters (density and scale-length, or virial mass and concentration)
against predictions from cosmological simulations.
N+09 assembled the PN-based results as well as a heterogeneous sample of
other mass results from the literature.  We reproduce this mass-concentration
plot in Fig.~\ref{fig:confb1}, with the theoretical prediction updated for
the WMAP5 cosmological parameters.

Although the mass profile uncertainties for any single galaxy are too large
to make definitive statements, when considering a handful of galaxies together,
a remarkable pattern begins to emerge.
The fast-rotator or ordinary ETGs (along with spiral galaxies)
appear to have low concentration haloes,
and the slow rotators to have high concentrations, with a possible zone of avoidance
in between, corresponding to the theoretical predictions.
With the shift to WMAP5 predictions, the low concentrations become {\it less} of a problem,
and the high concentrations {\it more}.

Adding NGC~4374 to the diagram confirms this picture with a PN-based
slow rotator analysis for the first time.
The NFW solution with a standard Kroupa IMF coincides with the high-concentration
region previously found for slow rotators using somewhat similar analyses.
However, the story changes with certain modifications to the models.
If the IMF is forced to Salpeter, less central DM is permitted and the
implied concentration decreases.
Alternatively, the high central DM content could be due to AC,
with the ``original'' concentration much lower, as illustrated by the modelling.
In either of these cases, the halo concentration becomes consistent with
$\Lambda$CDM predictions.

Selecting a ``heavy'' IMF or including AC may thus generally solve the concentration
crisis for slow rotators---but what about the fast rotators?
Although we have not explicitly modelled these galaxies with AC, some general trends may
be gleaned from the $\Lambda$CDM-based toy models of \citet{NRT10}.
Their Figs.~6 and 11 illustrate that for ETGs of all masses, AC is expected to
dramatically increase the fraction of DM found in the central \Re.
This implies that if AC were included in the models of the fast rotators of
Fig.~\ref{fig:confb1}, the halo concentrations which are currently on the margin
of consistency with theory would become problematically low.

An alternative scenario might be to adjust the IMFs of the fast rotators to be
{\it lighter} than Kroupa (Salpeter is incidentally too heavy in general for this
class of galaxies; cf. C+06).
This would allow for more central DM and conceivably increase the inferred
concentrations.

In order to move all the ``observed'' ETG halo concentrations into
reasonable agreement with the predictions, we arrive at the tentative solution that
(1) the slow rotators have Salpeter IMFs or AC, {\it and}
(2) the fast rotators have ultra-light IMFs or no AC.
If (1) and (2) are fulfilled, then there may still be a systematic concentration
offset between fast and slow rotators, but this would be small enough to be
plausibly explained by differing collapse redshifts.

This solution would present the very interesting possibility that the fast and
slow rotators are dramatically different in either their IMFs or their halo
contraction histories.
Systematic transitions in these properties have been suggested for various
reasons in the past, but they appear to go in the wrong direction.
In the modern ``downsizing'' picture of galaxy formation (e.g.,
\citealt{Nelan+05,Thomas+05,Cimatti+06,Pannella+06,Graves+07,Calura+08}),
the more massive galaxies like NGC~4374 would have on average
formed their stars earlier and more rapidly than in the more
ordinary ellipticals. As summarized by \citet[Sec. 4.4]{NRT10},
the IMF in these conditions is thought to have been if anything
{\it lighter} rather than heavier.

Also, as summarized in N+09, it is thought that AC could be counteracted by
rapid, clumpy and starbursty assembly histories, while AC classically implies
smooth, slow gaseous infall (see also \citealt{2010ApJ...712...88L}).
These conditions would suggest that the {\it spirals} should have stronger AC, and
galaxies like NGC~4374 should have weaker AC (a point also made by
\citealt{2008arXiv0808.0225C}).

\begin{figure}
\hspace{-0.5cm} \epsfig{file=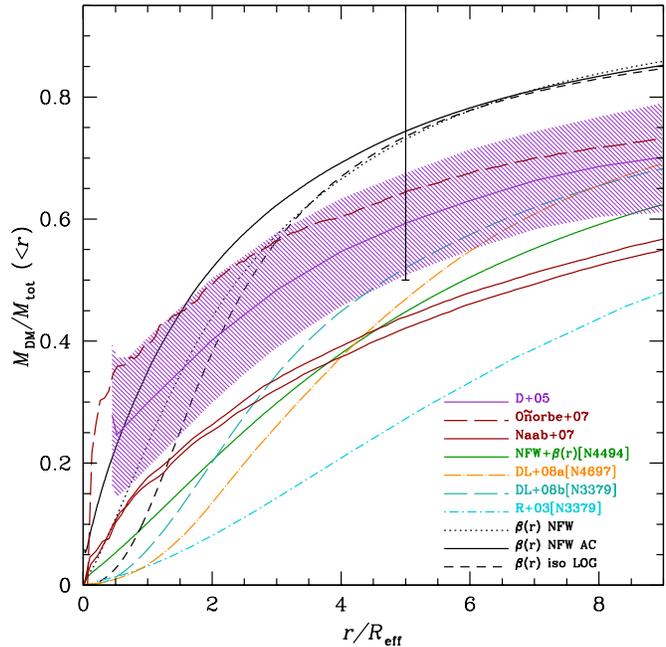,width=9.2cm}
\caption{Cumulative dark matter fraction as a function of radius.
Results for different observed and simulated galaxies are indicated with
different colours and linestyles as in the legend, with
the results for NGC 4374 in black. The errorbar marks the
typical error for the dark matter fraction of NGC~4374 at 5\Re.}
\label{fig:fDM}
\end{figure}

Returning to a less model-dependent view of the mass profiles, we plot the
cumulative DM fraction versus radius in Fig.~\ref{fig:fDM}, as also done in N+09.
The model inferences for NGC~4374 as well as some of the ordinary ETGs are plotted,
along with examples from galaxy formation simulations (both in a full cosmological
context and using ad-hoc mergers;
\citealt{2005Natur.437..707D,2007ApJ...658..710N,2007MNRAS.376...39O}).
Drawing attention to the 5~\Re\ reference value, we see that
the DM fraction for NGC~4374 of $\sim$~0.7--0.8 is significantly larger than
what was found so far for ordinary ellipticals ($\sim$~0.4--0.5),
and similar to what has been found for group- and cluster-central ellipticals
($\sim$~0.8--0.9 using X-ray rather than dynamical methods;
\citealt{2010arXiv1007.5322D}).
These results bracket the simulations values of $\sim$~0.5--0.6.

The DM fraction results within 1~\Re\
in Fig.~\ref{fig:fDM} based on detailed dynamical modelling
at first glance do not seem to square with other recent results
from the literature.  Various combinations of dynamical,
strong gravitational lensing, and stellar populations analyses have found
typical DM fractions within 1~\Re\ of $\sim$~0.4 for fainter ellipticals
and $\sim$~0.6 for brighter ones
\citep{NRT10,2009arXiv0911.2260S,2010ApJ...721L...1T,2010arXiv1007.2880A},
versus $\sim$~0.05 and $\sim$~0.3 here.

However, in the case of NGC~4374, the ambiguity in the \Re\ comes into play.
In NRT10, galaxies of the same stellar mass have $\Re\ \sim$~12~kpc on average,
or $\sim$~145$''$ at the distance of NGC~4374.  Using this \Re\ scale, we would
have a DM fraction of $\sim$~0.5, consistent with the literature.
As for the lower-luminosity ellipticals, NRT10 did find a fraction of galaxies
(particularly ones with older stars) to have DM fractions lower than $\sim$~0.1,
so the critical goal is to assemble a large sample of galaxies with
detailed dynamical models to establish the trends with good statistics.
Fig.~1 of \cite{2010arXiv1005.1289T} does suggest that these three galaxies may happen
to represent one extreme from a broad distribution of DM properties at
intermediate luminosities.
If this situation is true, the arguments above about halo concentration offsets
would no longer apply.

\section{Conclusions}\label{sec:concl}
We have presented a full Jeans analysis of the bright, slow-rotator
elliptical NGC~4374
based on the observations of $\sim$~450 PNe with the Planetary Nebula
Spectrograph. The PN line-of-sight velocities extend out to
$\sim$~5\Re. We have constructed spherical Jeans dynamical models of the
system:
a ``pseudo-inversion'' model and multi-component mass models with
and fourth-order moments constraints on the orbital anisotropy.

The two approaches return similar values of $M/L$ and
anisotropy (see Fig. \ref{fig:modelfitstwo} and Table
\ref{tab:jeanssumm}) and both imply that NGC~4374 is a very dark matter
dominated system with a near-isotropic orbital distribution in its halo.
Dynamical analyses of more ordinary ETGs have suggested radially-biased
anisotropy in their haloes as predicted by simulations (see Section~\ref{sec:anis_prof}).
The NGC~4374 result on the other hand would build on previous suggestions
that slow rotators have surprisingly isotropic haloes, which would
suggest a new scenario for building of the extended stellar envelopes of
these galaxies may be required \citep{2008ApJ...674..869H,2009AJ....137.4956R}.
However, in this case the anisotropy result is sensitive to the
assumptions about outlier velocities, and further investigation is required.

The mass profile results are on the other hand fairly insensitive to the outliers.
The high DM fraction inferred within $\sim$~5~\Re\ confirms the apparent
dichotomy in DM content between slow and fast rotators proposed by N+09
(see also \citealt{1994A&A...292..381B}; C+06; \citealt{2007IAUS..244..289N}; C+09),
and yields two important implications:
(1) the DM dichotomy is not a result of systematic differences in the mass
tracers used;
(2) it is not a simple difference of group-central versus satellite galaxies
since NGC~4374 does not appear to be at a group center (while the low-DM system
NGC~3379 {\it is}).

This apparent DM bimodality may mirror other transitions in ETG properties
at similar luminosity scales, such as the relations between size and mass
(e.g. \citealt{Shen2003,Tortora+09}),
size and surface brightness
(e.g.  \citealt{1992MNRAS.259..323C}),
luminosity and velocity dispersion
(\citealt{FJ76})
and the colour/population properties
(\citealt{2010MNRAS.407..144T}).

Given the limitations of the Jeans models and the stellar/dark mass degeneracy,
we are not able to distinguish between different DM radial profiles, including LOG,
NFW and NFW+AC haloes.
The LOG models prefer high stellar masses consistent with a Salpeter IMF,
NFW works with either Salpeter or Kroupa, and NFW+AC requires Kroupa.
The nominal NFW+Kroupa model implies a halo with a concentration that is
somewhat high, given WMAP5-based predictions.
Adopting either Salpeter IMF or AC brings the inferred concentration down to
the expected value.
Thus, considering that AC has commonly been considered the default expectation
in galaxy formation, we have finally found an ETG analyzed using PNe that is
naturally consistent with theoretical expectations for the DM halo.

Comparing the NFW halo parameters obtained for NGC~4374 as well as
for an assortment of other galaxies in the literature, we find
evidence for the slow rotators to have much higher halo
concentrations on average than the fast rotators.
We discuss some possible variations in IMF and AC which could explain
this difference, but there are also suggestions that the sample of
fast rotator galaxies is a statistical fluke.

Two primary avenues are needed to make further headway in pinning
down the properties of DM haloes in ETGs.
One is to carry out more detailed dynamical and stellar populations
modelling in an attempt to discern the DM profiles in detail.
The other is to expand the sample of galaxies studied, particularly at
intermediate luminosities ($M_B \sim -20$).
Work on both fronts is underway as part of the PN.S Elliptical Galaxy Survey.

\section*{Acknowledgments}
We would like to thank the anonymous referee for the fast report
and useful suggestions, and Isaac Newton Group staff on La Palma
for supporting the \PNS\ over the years. We also thank Crescenzo
Tortora for stimulating discussions.
AJR was supported by the National Science Foundation Grants
AST-0507729, AST-0808099 and AST-0909237, and by the FONDAP Center
for Astrophysics CONICYT 15010003.
This research has made use of the NASA/IPAC Extragalactic
Database (NED) which is operated by the Jet Propulsion Laboratory,
California Institute of Technology, under contract with the
National Aeronautics and Space Administration. We acknowledge the
usage of the HyperLeda database (http://leda.univ-lyon1.fr).

\appendix

\section[]{Alternative outlier selection and dynamical implications}\label{app:appA}

As discussed at the beginning of Section~\ref{sec:sample}, a
handful of ``outlier'' PNe were rejected from the overall sample
using a 3~$\sigma$ ``friendless'' analysis.
Although relatively few in number, the inclusion or exclusion of
these objects in our dynamical analyses could have a large impact
on the conclusions, which we consider here in more detail.

Fig.~\ref{fig:outli} showed the six outliers identified through
this process.  Two of them are extreme outliers and can be
securely rejected, but  the other four are only barely excluded at
3~$\sigma$. This is a concern since in a data set of 450 objects
with a Gaussian velocity distribution, there should on average be
one random object found past 3~$\sigma$, and if a non-Gaussian
distribution is allowed, then many more would be
possible\footnote{We have checked how the outlier velocities
compare to the local escape velocities in our best-fit NFW halo
(e.g. the NFW+AC+$\beta(r)$ in Table \ref{tab:jeanssumm}), which
turns out to be $\sim$~1250~\kms relative to the systemic
velocity. The two most extreme outliers would in this case not be
bound to NGC~4374, but the other four could be.}.

We are not at a complete impasse since we notice that all
the outliers have negative velocities relative to NGC~4374, which
is not likely to be just a chance occurrence\footnote{This asymmetry
does not appear to be caused by an error in the adopted value of
$v_{\rm sys}$, as the peak of the LOSVD coincides with our
self-consistent $v_{\rm sys}$, which is in turn very close to the NED value.}.

\begin{figure}
\centering \hspace{-0.5cm}
\epsfig{file=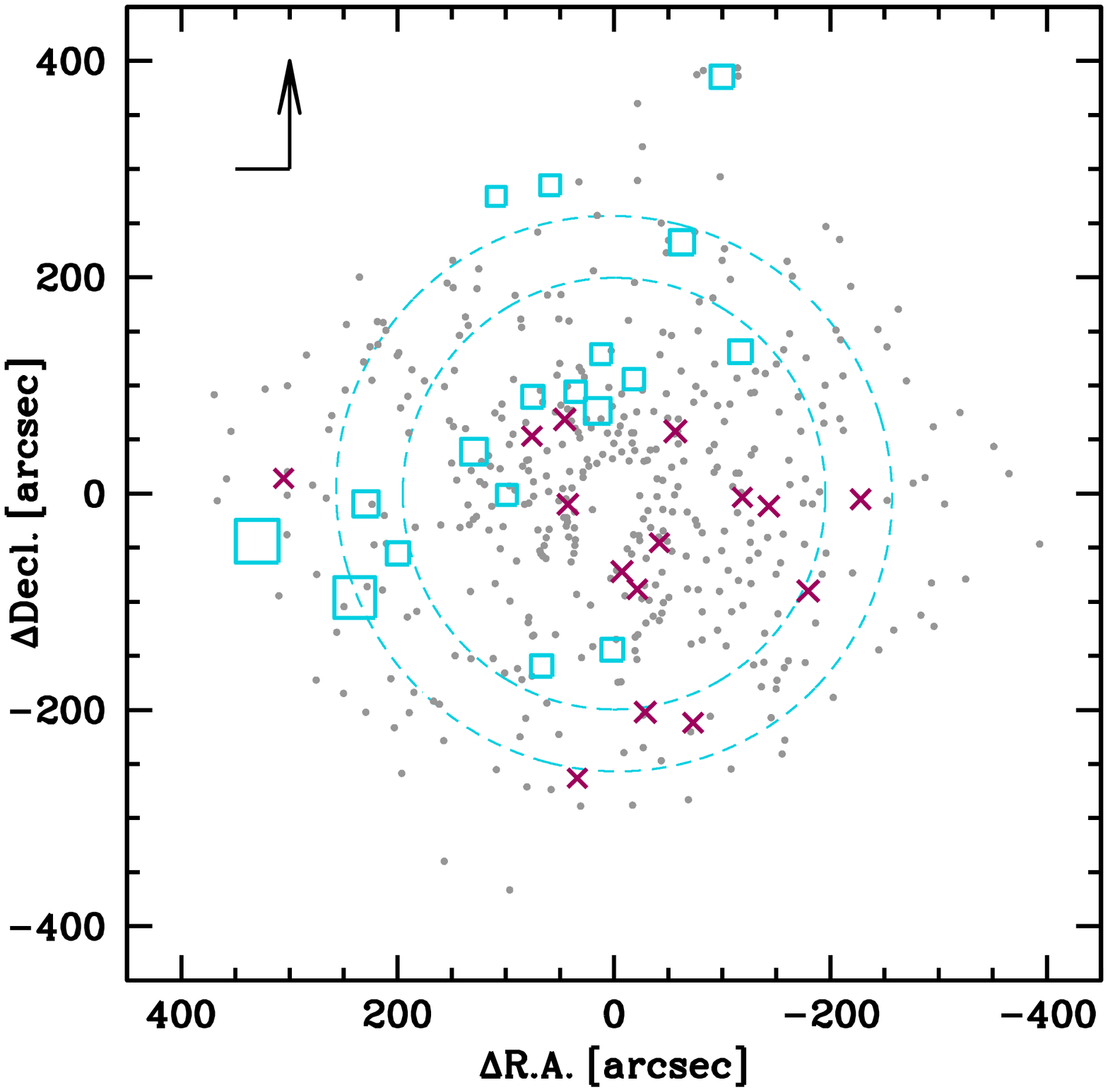,width=8.7cm}
\caption{Diagram of potential velocity outlier PNe.
The 2D positions on the sky are shown relative to the center of
NGC~4374, with the stellar isophotes at $R_m =197''$ and 257$''$
show as dashed ellipses.
Squares represent approaching velocities and crosses are receding velocities,
with symbol size proportional to relative velocity amplitude.
The $\sim$~30 most extreme velocity PNe are shown along with the
candidate outliers, to illustrate the normal velocity field of NGC~4374.
The remaining PNe are shown as small grey points.
The North and East directions are
shown in the top-left corner. }\label{fig:figA1}
\end{figure}

We look at the situation in two-dimensions in Fig.~\ref{fig:figA1},
focusing on the most outlying velocities.
It turns out that the four most extreme velocities lie on an axis
to the East of the galaxy's center, which is also the direction of
the nearby giant elliptical NGC~4406 (M86) found $\sim 1000''$ away.
A similar pattern has been found in the globular cluster system of NGC~4374
(B. Kumar et al., in prep).

The $-1300$~\kms\ relative systemic velocity of NGC~4406 ($v_{\rm
rel}$) provides a handy explanation for the low-velocity
outliers---whether these objects simply belong to the NGC~4406 PN
system seen in projection, or are part of an interaction region
between the two galaxies (\citealt{arn96}).

Although a full analysis for such an interaction scenario is
outside the scope of this paper, we can quantify the effect of a
fly-by encounter between the two galaxies using the {\it impulse
approximation} to estimate the energy injection into the outer
galaxy envelope
(see e.g.  \citealt{nap02}).
We calculate an upper limit to this energy
by assuming a tangential encounter with an impact parameter of
$b=1000''$:
\begin{equation}
\Delta E=\frac{4GM_1 M_2^2}{3 b^4 v_{\rm rel}^2}\overline{r^2} ,
\end{equation}
where $G$ is the gravitational constant, $M_1\sim6\times
10^{12}\Msun$ (e.g., from the ``NFW+iso2'' model) is the mass of
the perturbed system (NGC 4374) which has been calculated within
the impact parameter $b$, $M_2=0.5 \times M_1$ at the same radius, and the
mean square radius of NGC 4374, $\overline{r^2}$ is taken as
equivalent to the square of characteristic scale of the dark
matter halo ($\sim 6.4\times 10^3$ kpc$^2$).

The resulting energy change is $\Delta
E=9.2\times 10^{16}$ \Msun\ km$^2$ s$^{-2}$ which provides a
heating contribution to the dispersion of
$\sigma_{\rm heat}=2\Delta E/(3 M_{\rm shell})$, where $M_{\rm shell}$
is the mass of the galaxy shell which has experienced the energy
transfer. Taking this shell in the radial range of $200''$--$1000''$
(i.e. $\gsim 3\Re$), we find $\sigma_{\rm heat}\sim 100\kms$.

This extra heating term could handily explain the higher
dispersion on the low velocity side as implied by the ``outliers''
in Fig.~\ref{fig:outli}. In this scenario, the close passage
between the galaxies would have heated the Eastern side of
NGC~4374, with this event happening less than one-crossing time
ago so that the asymmetry is preserved. Removing the four
``outliers'' would then restore the observed kinematics of the
system to the approximate pre-interaction state, suitable for
equilibrium dynamical analyses.

The interaction calculation above has been done under the
assumption of the closest encounter (and the highest energetic)
allowed by the observed geometry. Any other less favourable
configuration would produce a smaller energy transfer and a more
local effect of the encounter. In this case, the four Eastern
low-velocity objects are likely to be true outliers, and the
remaining two outliers to the North are less certain, and could be
part of the normal velocity distribution of NGC~4374.

\begin{figure*}
\hspace{-0.7cm}
\begin{minipage}[b]{0.35\linewidth}
\includegraphics[scale=0.57]{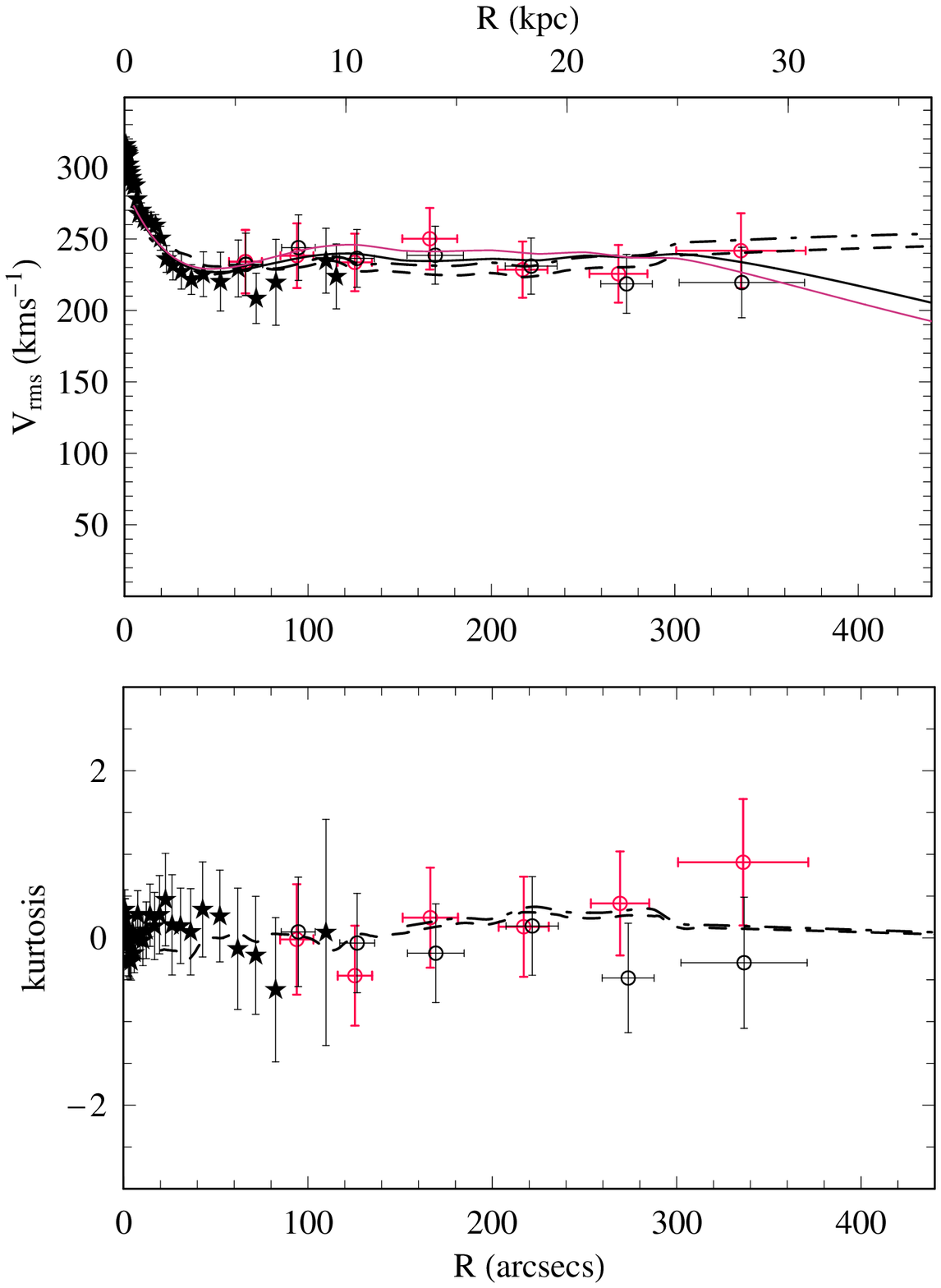}
\end{minipage}
\hspace{3cm}
\begin{minipage}[t]{0.5\linewidth}
\vspace{-10.82cm}
\includegraphics[scale=0.69]{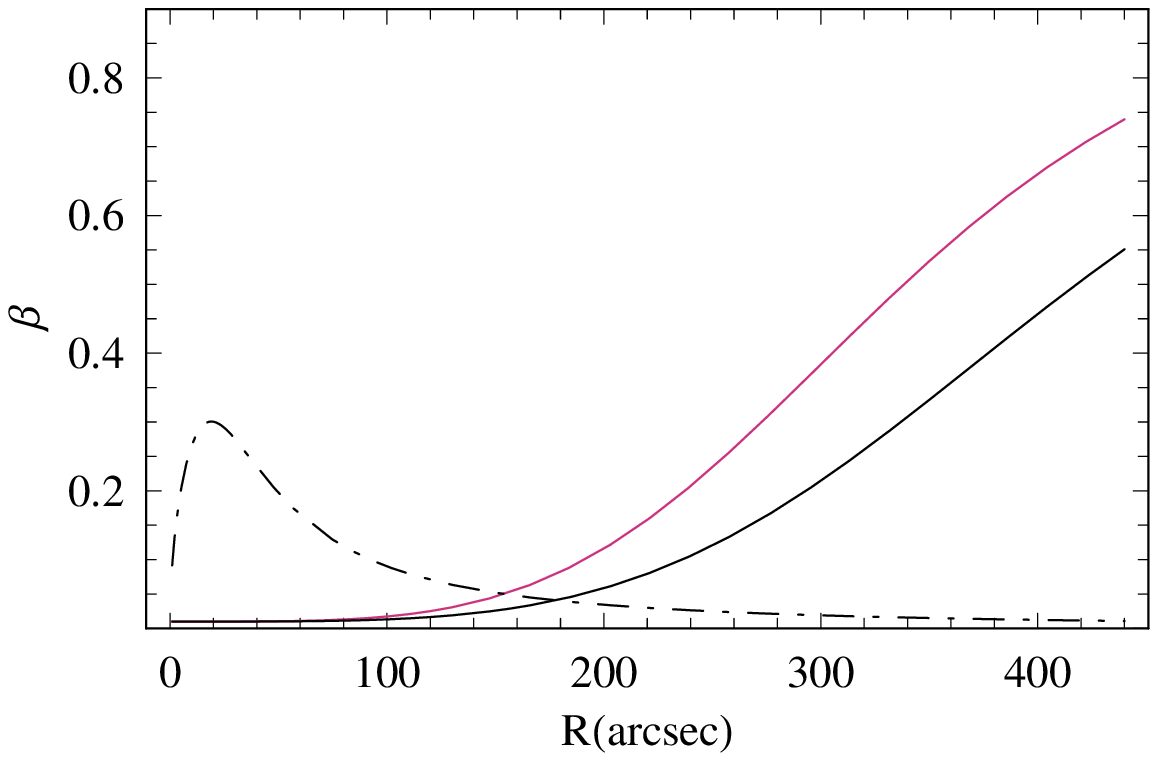}
\includegraphics[scale=0.75]{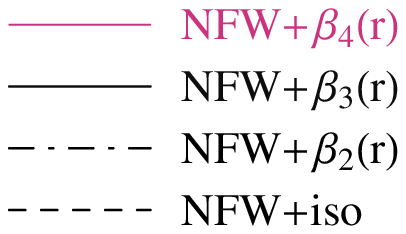}
\end{minipage}

\caption{Effect of the outliers selection. {\it Top:} the velocity
dispersion profile of the outliers selection with the friendless
algorithm, adopted in the paper (black) is compared with the
dispersion obtained by including the two uncertain outliers along
the $3\sigma$ borderline (red). {\it Bottom:} the same for the
kurtosis profile. Overplotted there are the isotropic model
obtained for both profiles, and a best fit model with a steeply
increasing $\beta(r)$ profile as suggested by the kurtosis
including the two uncertain outliers, as in the top-right panel.
Models are as in the legend. See text for details.}\label{fig:figA2}

\end{figure*}

If those two objects are kept in the final sample then the velocity
dispersion and kurtosis profiles are somewhat changed in the outer
regions, as shown in Fig. \ref{fig:figA2}.
The dispersion profile becomes slightly flatter
(slope $-0.03\pm0.07$ instead of the $-0.07$ found in
\S\ref{sec:dispsec}).
The kurtosis profile rises at large radii, where if we were to again use
the equation B10 approximation from N+09,
we would infer a higher radial anisotropy ($\beta\sim +0.4$
instead of $\sim -0.1$).

Carrying out some dynamical models as in the main Sections,
we show in Fig.~\ref{fig:figA2}
the results for the isotropic and $\beta(r)$ NFW mass models.
We find best-fitting halo parameters of $\rho_s=0.0019, 0.0030$
$\Msun$~pc$^{-3}$ and $r_s=110, 87$~kpc respectively,
corresponding to $c\vir=8^{+5}_{-2}, 10^{+6}_{-4}$ and $\log
M\vir\sim13.6\Msun$. These parameters are very similar to those
found using our default outlier selection (see Figs.
\ref{fig:fig7} and \ref{fig:fig8}), although the $\chi^2$ fits are poorer.

We also try out a more strongly varying $\beta(r)$ function
motivated by the higher kurtosis, with high radial anisotropy at
larger radii as illustrated by the right-hand panel of
Fig.~\ref{fig:figA2} and named $\beta_3$ and $\beta_4$ some fixed
profiles which bracket the tentative anisotropy value in the
latest radial bin estimated as above ($\sim+0.4$). In this case
the fit is performed on the dispersion curve only.

The quality of the corresponding dynamical model fit (top-left
panel) is similar to the previous case, but the best-fit dark
matter halo turns out to be almost identical for the two
anisotropy profiles and have a higher halo concentration and small
virial mass, slightly off the WMAP5--$\Lambda$CDM predictions:
$\rho_s=0.006$ $\Msun$~pc$^{-3}$ and $r_s=46$ kpc, corresponding
to $c\vir=13^{+9}_{-6}$ and $\log M\vir\sim13.1\pm0.1\Msun$, whith
errors including the variance of the assumed $\beta$ profiles.
In both these strongly radial models, the velocity dispersion
bends quite significantly outside the last dispersion bin, which
is a prediction that should be tested with more extended data.

We thus find that the impact of the outlier ambiguity is confined
to the anisotropy conclusions, with highly radial halo orbits
suggested by the kurtosis but hardly matched by the dispersion
profile which is flatter when the two uncertain outliers are
included.
The mass profile inferences are presumably unaffected because of
the pinch-point phenomenon, whereby the projected dispersion is
only weakly dependent on anisotropy in certain regions of the
galaxy. Further observations of PNe at larger radii (see e.g.
\citealt{2004ApJ...614L..33A}), particularly on the West side of
the galaxy, could clarify the situation by more strongly
constraining the dispersion and kurtosis profiles past the pinch
point.

\bibliography{napolitano_pns_v2}

\end{document}